\documentclass[12pt]{iopart}
\usepackage{iopams}  

\eqnobysec

\begin{document}

\newtheorem{proposition}{Proposition}[section]

\renewcommand\ge{\geqslant} 
\renewcommand\le{\leqslant} 

\newcommand\mymatrix[1]{\mathsf{#1}}
\newcommand\myset[1]{\mathrm{#1}}
\newcommand\mysetsize[1]{|\myset{#1}|}
\newcommand\myShift[2][T]{\mathbb{#1}_{#2}}
\newcommand\myShifted[3][T]{\mathbb{#1}_{#2}{#3}}
\newcommand\myBarShift[2][T]{\overline{\mathbb{#1}}_{#2}}
\newcommand\myBarShiftInv[2][T]{\overline{\mathbb{#1}}_{#2}^{-1}}
\newcommand\myBarShifted[3][T]{\overline{\mathbb{#1}}_{#2}{#3}}
\newcommand\myshifted[2]{{#2}[{#1}]}
\newcommand\myalh[2]{\if ?#2? #1\, \else \myshifted{#2}{#1} \fi}
\newcommand\myGamma[3]{\if ?#2? \Gamma_{#1}(#3)\, \else \Gamma_{#1}(#2,#3)\, \fi}
\newcommand\myHatGamma[3]{\if ?#2? \hat\Gamma_{#1}(#3) \else \hat\Gamma_{#1}(#2,#3) \fi}
\newcommand\mySum[2]{\sum_{#1 \in \myset{#2}}}
\newcommand\myHirota[3]{D_{#1}\, #2\cdot #3}
\newcommand\myBarHirota[3]{\bar{D}_{#1}\, #2\cdot #3}
\newcommand\myf{f}


\title{Soliton Fay identities. II. Bright soliton case.}
\author{V.E. Vekslerchik}
\address{
  Usikov Institute for Radiophysics and Electronics \\
  12, Proskura st., Kharkov, 61085, Ukraine 
}
\ead{vekslerchik@yahoo.com}
\submitto{\JPA}

\begin{abstract}
We present a set of bilinear matrix identities that generalize the ones that 
have been used to construct the bright soliton solutions for various models. 
As an example of an application of these identities, we present a simple 
derivation of the $N$-bright soliton solutions for the Ablowitz-Ladik 
hierarchy.
\end{abstract}

\section{Introduction.}

This paper is a continuation of the work initiated in \cite{V14} where we 
presented the set of the soliton Fay identities.
The main aim of \cite{V14} was to summarize the results of a number of works 
devoted to the soliton solutions for various integrable models.
We studied some properties of the matrices that had been used in 
\cite{V05,PV11,V11,V12a,V13a,V13b} to construct the dark solitons 
and presented some identities that are satisfied by these matrices. 
These identities, which may be viewed as soliton analogues of the Fay 
identities \cite{Fay,Mumford}, can be used as a starting point 
for the so-called direct approach to the integrable equations.

The classical Fay identities \cite{Fay,Mumford} have been derived for the 
theta-functions associated with compact Riemann surfaces of the finite genus. 
It is a known fact that the so-called finite-gap solutions for integrable 
equations (which are built of these theta-functions) can be transformed 
into soliton ones by degenerating the corresponding Riemann surface 
(see \cite{BBEIM94,GH03} and references therein).
Thus, the soliton identities of \cite{V14} and of this paper can be, 
in principle, obtained by limiting procedure form the classical Fay identities. 
However, we do not use such approach, since our identities can be derived by 
elementary calculations without invoking the much more difficult 
algebro-geometric constructions of \cite{Fay}. 
In other words, we use the term `Fay identity' to refer to bilinear 
identities that are derived from the properties of the 
involved objects (theta-functions for classical Fay identities, and soliton 
matrices for our  case).

Solutions that can be obtained from the Fay identities presented in \cite{V14}, 
in most of the cases, correspond to the so-called `dark' solitons which obey 
nonvanishing finite-density (\textit{i.e.} with constant or bounded modulus) 
boundary conditions. 
In this paper we present the results that lead to the so-called `bright' 
solitons (localized structures, vanishing at the infinity). 
It was noted in \cite{V14}, and we would like to repeat it here, that 
usage of the terms `dark' in \cite{V14} and `bright' in this paper is somewhat 
dubious because we, first, do not discuss the boundary value
problems. Second, as it can be seen from examples in \cite{V14}, the 
behaviour of solutions can differ from the one of dark solitons and, probably, 
not all solutions that can be obtained from the identities presented here are 
indeed bright solitons. 
We use the words `dark' and `bright' to distinguish the two types of, 
so to say, architecture of the soliton solutions for already known, classical, 
integrable models. 
It is easy to see that the determinants discussed in \cite{V14} do not vanish 
in the asymptotical regions: their `natural' limiting values are $1$ or $\infty$, 
so they are not the best candidates for the role of the vanishing at the 
infinity tau-functions, which one needs to build bright solitons. 
An inspection of already known bright-soliton solutions shows that usually 
some of the corresponding tau-functions have different structure. 
Namely these objects, not only the determinants similar to the ones of 
\cite{V14}, are considered next, and that is the reason to use the term 
`bright' in the title of this paper.

As was mentioned in \cite{V14}, the dark solitons, if considered in the 
framework of the inverse scattering transform, are, in some sense, more 
`difficult' objects than the bright ones because to derive them 
one has to study the spectral problems with nonvanishing at the infinity 
potentials, which complicates the involved mathematics. 
However, in the framework of the direct approach the situation seems to be 
inverted. Indeed, looking at the results presented next, a reader can note 
that sometimes they are more cumbersome than their analogues from \cite{V14}. 

As in \cite{V14}, we start with introducing in section \ref{sec-mat} a class 
of `soliton' matrices. These matrices are used to construct the tau-functions 
for which we deduce, by simple algebra, a set of \textit{bilinear} identities 
(section \ref{sec-fay}) and present their alternative formulations 
(section \ref{sec-miwa}) and their differential versions 
(section \ref{sec-diff}). In section \ref{sec-app}, we discuss an application 
of the obtained identities. Contrary to the previous paper \cite{V14}, we do 
not present here new integrable models. Instead, we discuss in detail the 
case of the well-known Ablowitz-Ladik (AL) system \cite{AL75,AL76,AS,APT} 
pursuing two goals. 
First we want to demonstrate how to solve some technical problems 
that one can face when applying general formulae to a particular equation. 
Another motivation is to fill a certain gap: although the AL model is one 
the first models shown to be integrable, one can hardly find in the literature 
a compact expression, similar to the one derived in section \ref{sec-app}, for 
the $N$-bright soliton solution for the whole AL \textit{hierarchy} (ALH).

\section{Soliton matrices and tau-functions. \label{sec-mat}}

In this section we introduce the main objects of our study: soliton matrices, 
shifts, and tau-functions, and discuss some of their properties that we need to 
derive the Fay identities.

\subsection{Matrices and shifts.}
As in \cite{V14}, the Fay identities, that we present next, are formulated in 
terms of the so-called `almost-intertwining' matrices \cite{KG01} that satisfy 
the `rank one condition' \cite{GK02,GK06}, which is a particular case of 
the Sylvester equation (see \cite{NAH09,DM10a,DM10b,ZZ13,XZZ14}). 

Contrary to the dark-soliton case \cite{V14}, we need two families of 
$N \times N$ matrices, $\mymatrix{A}$ and $\mymatrix{B}$, that solve 
`conjugated' equations, 
\begin{equation}
  \begin{array}{lcl}
  \mymatrix{L} \mymatrix{A} - \mymatrix{A} \mymatrix{R} 
  & = & 
  | \,\alpha\, \rangle \langle a |, 
  \\
  \mymatrix{R} \mymatrix{B} - \mymatrix{B} \mymatrix{L} 
  & = & 
  | \,\beta\, \rangle \langle b |. 
  \end{array}
\label{eq-Sylvester} 
\end{equation}
Here, 
$\mymatrix{L}$ and $\mymatrix{R}$ are diagonal constant matrices, 
$| \,\alpha\, \rangle$ and $| \,\beta\, \rangle$ are constant $N$-component 
columns, 
$\langle a |$ and $\langle b |$ are $N$-component rows that depend on the 
coordinates describing the model we are dealing with.

The second element of our construction are shifts $\myShift{\xi}$. 
From the viewpoint of applications, these shifts can have various origins. 
They, for example, may present the lattice translations or to be a tool to 
model the so-called Miwa shifts, if one deals with the continuous problems.
However, for our current purposes their `physical sense' is not important and 
we \emph{define} them by their action upon vectors 
and matrices introduced previously: 
\begin{equation}
  \begin{array}{lcl}
  \myShifted{\xi}{\langle a |} 
  & = & 
  \langle a | \left(\mymatrix{R} - \xi\right)^{-1}, 
  \\[2mm]
  \myShifted{\xi}{\langle b |} 
  & = & 
  \langle b | \left(\mymatrix{L} - \xi\right) 
  \end{array} 
\label{def-shits-ab}
\end{equation}
which, in the general case (when the elements of $\mymatrix{L}$ are distinct 
from the elements of $\mymatrix{R}$), leads to 
\begin{equation}
  \begin{array}{lcl}
  \myShifted{\xi}{\mymatrix{A}} 
  & = & 
  \mymatrix{A} \left(\mymatrix{R} - \xi\right)^{-1}, 
  \\[2mm]
  \myShifted{\xi}{\mymatrix{B}} 
  & = & 
  \mymatrix{B} \left(\mymatrix{L} - \xi\right). 
  \end{array} 
\label{def-shits-AB}
\end{equation}
The action of these shifts on other objects that appear in this paper 
(such as determinants, tau-functions \textit{etc}) can be easily `prolonged' 
from \eref{def-shits-ab} and \eref{def-shits-AB} by the simple rule: 
$\myShift{\xi}$ is the lowest-level operation preceding all algebraic ones, 
i.e. 
$\myShift{\xi}
  f( 
    \langle a |, 
    \langle b |, 
    \mymatrix{A}, 
    \mymatrix{B}, ... 
  ) 
  = 
  f( 
    \myShifted{\xi}{\langle a |}, 
    \myShifted{\xi}{\langle b |}, 
    \myShifted{\xi}{\mymatrix{A}}, 
    \myShifted{\xi}{\mymatrix{B}}, ... 
  ) 
$ for any function $f$.
In the general case, these shits can be inverted by inverting the matrices 
appearing in the right-hand sides of 
\eref{def-shits-ab} and \eref{def-shits-AB}: 
$
  \myShift{\xi}^{-1} \langle a | 
  =  
  \langle a | \left(\mymatrix{R} - \xi\right) 
$, 
$
  \myShift{\xi}^{-1} \langle b | 
  = 
  \langle b | \left(\mymatrix{L} - \xi\right)^{-1}
$ 
and similar formulae for $\mymatrix{A}$ and $\mymatrix{B}$, 
and then extending the action of $\myShift{\xi}^{-1}$ 
by
$\myShift{\xi}^{-1}
  f( 
    \langle a |, 
    \langle b |, 
    \mymatrix{A}, 
    \mymatrix{B}, ... 
  ) 
  = 
  f( 
    \myShift{\xi}^{-1}\langle a |, \allowbreak
    \myShift{\xi}^{-1}\langle b |, 
    \myShift{\xi}^{-1}\mymatrix{A}, 
    \myShift{\xi}^{-1}\mymatrix{B}, ... 
  ) 
$.

As in \cite{V14}, we use the set notation to 
write a superposition of the shifts $\myShifted{\xi}{}$: 
for any set of $\xi$, 
$ \myset{X} = \{ \xi_{1}, ..., \xi_{n} \}$, the symbol  
$\myShift{\myset{X}}$ stands for 
\begin{equation}
  \myShift{\myset{X}} 
  = 
  \prod_{\xi \in \myset{X} } \myShift{\xi}. 
\end{equation}
There is no restriction on the number of appearances of any symbol in a set: 
if  
$\myset{X} 
  = 
  \{ \overbrace{\xi, ... , \xi}^{n}, ...  \},
$
then 
$
  \myShift{\myset{X}} 
  = 
  \myShift{\xi}^{n}... 
$
In what follows we use $|...|$ for the number of elements of a set 
(if $\myset{X} = \{ \xi_{1}, ..., \xi_{n} \}$, then 
$ \mysetsize{X} = n$, even if some of $\xi_{i}$ coincide). 
We denote by $\myset{X}\myset{Y}$ the sum of sets 
(if any symbol $\xi$ appears $m$ times in $\myset{X}$ and $n$ times in 
$\myset{Y}$, then there are $m+n$ symbols $\xi$ in the resulting set)
and write $\myset{X}/\xi$ ($\xi \in \myset{X})$ for the 
set $\myset{X}$ with the element $\xi$ being excluded.

Sometimes, to make the formulae more readable, we avoid the 
$\myShifted{}{}$-notation and write the shifted quantities as follows:
\begin{equation}
  \myshifted{\myset{X}}{\mymatrix{A}} 
  = 
  \myShifted{\myset{X}}{\mymatrix{A}}. 
\end{equation}

\subsection{Tau-functions}
Finally, we introduce the triplet of tau-functions, 
$\tau$, $\sigma$ and $\rho$, defined by 
\begin{equation}
  \tau 
  = 
  \det \left| \mymatrix{1} + \mymatrix{A}\mymatrix{B} \right|
  = 
  \det \left| \mymatrix{1} + \mymatrix{B}\mymatrix{A} \right|
\label{def-tau} 
\end{equation}
and 
\begin{equation}
  \begin{array}{lcl}
  \sigma & = & \tau \langle a | \mymatrix{F} | \beta \rangle, 
  \\[2mm] 
  \rho & = & \tau \langle b | \mymatrix{G} | \alpha \rangle  
  \end{array} 
\label{def-sigmarho} 
\end{equation}
where matrices $\mymatrix{F}$ and $\mymatrix{G}$ are given by 
\begin{equation}
  \begin{array}{lcl}
  \mymatrix{F} 
  & = & 
  ( \mymatrix{1} + \mymatrix{B}\mymatrix{A} )^{-1}, 
  \\ 
  \mymatrix{G} 
  & = & 
  ( \mymatrix{1} + \mymatrix{A}\mymatrix{B} )^{-1}. 
  \end{array} 
\label{def-FG}
\end{equation}

\subsection{Matrix identities.}

One can deduce from the fact that matrices $\mymatrix{A}$ and $\mymatrix{B}$ 
solve \eref{eq-Sylvester} the following identities:
\begin{eqnarray} 
  \myshifted{\zeta}{\tau} / \tau 
  & = & 
  1 + 
  \langle \myshifted{\zeta}{a} | \mymatrix{F} \mymatrix{B} | \alpha \rangle  
  = 
  1 - 
  \langle b | \mymatrix{G} \myshifted{\zeta}{\mymatrix{A}} | \beta \rangle, 
\label{shifted-tau-p} 
\\ 
  \myshifted{\bar\zeta}{\tau} / \tau 
  & = & 
  1 - 
  \langle a | \mymatrix{F} \myshifted{\bar\zeta}{\mymatrix{B}} 
  | \alpha \rangle  
   = 
  1 + 
  \langle \myshifted{\bar\zeta}{b} | \mymatrix{G}\mymatrix{A} | \beta \rangle 
\label{shifted-tau-n} 
\end{eqnarray} 
and
\begin{eqnarray}
  \myshifted{\zeta}\sigma / \tau 
  & = & 
  \langle \myshifted{\zeta}{a} | \mymatrix{F} | \beta \rangle  
  = 
  \langle a | \mymatrix{F} (\mymatrix{R} - \zeta)^{-1} | \beta \rangle, 
\label{shifted-sigma} 
\\
  \myshifted{\bar\zeta}{\rho} / \tau 
  & = & 
  \langle \myshifted{\bar\zeta}{b} | \mymatrix{G} | \alpha \rangle  
  = 
  \langle b | \mymatrix{G} (\mymatrix{L} - \zeta)^{-1} | \alpha \rangle  
\label{shifted-rho} 
\end{eqnarray}
where overlined letters indicate the inverse shifts,
\begin{equation}
  \myshifted{\bar\zeta}{\omega} 
  = 
  \myShift{\zeta}^{-1} \omega.
\end{equation}
One can prove \eref{shifted-tau-p} and \eref{shifted-tau-n}  following the 
lines described in Appendix A of \cite{V14} (we do not repeat these 
calculations here). A proof of \eref{shifted-sigma} is presented in 
\ref{proof-A}.

A bit more complicated calculations lead to the following set of 
$
\left( \myShift{\zeta}{\omega'} \right)
\left( \myShift{\zeta}^{-1}{\omega''} \right)
$
identities:
\begin{eqnarray}
\label{innerTT}
  \frac{ \myshifted{\myset{X}\bar\zeta}{\tau} 
         \myshifted{\myset{Y}\zeta}{\tau} }
       { \myshifted{\myset{X}}{\tau} 
         \myshifted{\myset{Y}}{\tau} }
  & = & 
  1 + 
  \langle \myshifted{\myset{X}}{a} | 
  \myshifted{\myset{X}}{\mymatrix{F}}
  \mathcal{K}_{\tau\tau}(\zeta,\myset{X},\myset{Y})
  \myshifted{\myset{Y}}{\mymatrix{G}}  
  | \alpha \rangle, 
\\
  \frac{ \myshifted{\myset{X}\zeta}{\sigma} 
         \myshifted{\myset{Y}\bar\zeta}{\rho} }
       { \myshifted{\myset{X}}{\tau} 
         \myshifted{\myset{Y}}{\tau} }
  & = &
  \langle \myshifted{\myset{X}}{a} | 
  \myshifted{\myset{X}}{\mymatrix{F}}
  \mathcal{K}_{\sigma\rho}(\zeta,\myset{Y})
  \myshifted{\myset{Y}}{\mymatrix{G}}  
  | \alpha \rangle,  
\\
  \frac{ \myshifted{\myset{X}\zeta}{\sigma} 
         \myshifted{\myset{Y}\bar\zeta}{\tau} }
       { \myshifted{\myset{X}}{\tau} 
         \myshifted{\myset{Y}}{\tau} }
  & = & 
  \langle \myshifted{\myset{X}}{a} | 
  \myshifted{\myset{X}}{\mymatrix{F}}
  \mathcal{K}_{\sigma\tau}(\zeta,\myset{Y})
  \myshifted{\myset{Y}}{\mymatrix{F}}  
  | \beta \rangle,  
\label{innerST}
\\
  \frac{ \myshifted{\myset{X}\bar\zeta}{\rho} 
         \myshifted{\myset{Y}\zeta}{\tau} }
       { \myshifted{\myset{X}}{\tau} 
         \myshifted{\myset{Y}}{\tau} }
  & = & 
  \langle \myshifted{\myset{X}}{b} | 
  \myshifted{\myset{X}}{\mymatrix{G}}
  \mathcal{K}_{\rho\tau}(\zeta,\myset{Y})
  \myshifted{\myset{Y}}{\mymatrix{G}}  
  | \alpha \rangle  
\end{eqnarray}
where 
\begin{eqnarray}
  \mathcal{K}_{\tau\tau}(\zeta,\myset{X},\myset{Y})
  & = & 
  \Delta(\mymatrix{R}, \myset{X}, \myset{Y} ) 
  (\mymatrix{R} - \zeta)^{-1} 
  \myshifted{\myset{Y}}{\mymatrix{B} } 
  - 
  \myshifted{\myset{X}}{\mymatrix{B} }
  (\mymatrix{L} - \zeta )^{-1},
\\
  \mathcal{K}_{\sigma\rho}(\zeta,\myset{Y})
  & = & 
  \myshifted{\myset{Y}}{\mymatrix{B} } 
  (\mymatrix{L} - \zeta)^{-1} 
  - 
  (\mymatrix{R} - \zeta)^{-1} 
  \myshifted{\myset{Y}}{\mymatrix{B} }, 
\\ 
  \mathcal{K}_{\sigma\tau}(\zeta,\myset{Y})
  & = & 
    (\mymatrix{R} - \zeta)^{-1}  
    + 
    \myshifted{\myset{Y}}{\mymatrix{B} } 
    (\mymatrix{L} - \zeta)^{-1} 
    \myshifted{\myset{Y}}{\mymatrix{A} },
\\ 
  \mathcal{K}_{\rho\tau}(\zeta,\myset{Y})
  & = & 
    (\mymatrix{L} - \zeta)^{-1}  
    + 
    \myshifted{\myset{Y}}{\mymatrix{A} } 
    (\mymatrix{R} - \zeta)^{-1} 
    \myshifted{\myset{Y}}{\mymatrix{B} }
\label{defKRT}
\end{eqnarray}
and
\begin{equation}
  \Delta(\mymatrix{R}, \myset{X}, \myset{Y} ) 
  = 
  \prod_{\xi \in \myset{X}}
  (\mymatrix{R} - \xi)
  \prod_{\eta \in \myset{Y}}
  (\mymatrix{R} - \eta)^{-1}. 
\end{equation}
Identity \eref{innerTT} and its derivation is similar to the ones presented in 
\cite{V14} (we do not repeat the calculations here). 
We prove \eref{innerST} in \ref{proof-B} and note that the remaining 
identities can be obtained in the same manner. 
In the next section we use this result to derive the bright-soliton Fay 
identities.

\section{Fay identities. \label{sec-fay}}
%
\subsection{Derivation of the main identities.}
Now we use the presentation of 
$
\left( \myShift{\zeta}{\omega'} \right)
\left( \myShift{\zeta}^{-1}{\omega''} \right)
$
as meromorphic functions of $\zeta$ to obtain `external' identities which can 
be written in terms of the shifts, without exposing the inner structure of the 
tau-functions.

This part of calculations is rather simple. All we need is the elementary 
formula for a quotient of polynomials: if 
\begin{equation}
  \Delta(t,\myset{X}) = 
  \prod_{ \xi \in \myset{X} }(t-\xi) 
\end{equation}
and $\mysetsize{Y} < \mysetsize{X}$, then 
\begin{equation}
  \frac{ \Delta(t,\myset{Y}) }{ \Delta(t,\myset{X}) }
  = 
  \sum_{ \xi \in \myset{X} } 
  \frac{ \myGamma{\xi}{\myset{Y}}{\myset{X}} }{ t - \xi }
\end{equation}
where
\begin{equation}
  \myGamma{\xi}{\myset{Y}}{\myset{X}} 
  = 
  \frac{ \Delta( \xi, \myset{Y}) }{ \Delta( \xi, \myset{X}/\xi) }.
\end{equation}
Multiplying identities \eref{innerTT}--\eref{defKRT} by 
$\myGamma{\zeta}{...}{\myset{X'}}$ or 
$\myGamma{\zeta}{...}{\myset{Y'}}$, 
summarizing over $\myset{X'}\subset\myset{X}$ or $\myset{Y'}\subset\myset{Y}$ 
and noting that 
\begin{equation}
  \sum_{ \xi \in \myset{X} } 
  \myGamma{\xi}{\myset{Y}}{\myset{X}} 
  = \delta_{\mysetsize{X},\mysetsize{Y}+1} 
\qquad
  (\mysetsize{X} > \mysetsize{Y}) 
\end{equation}
one arrives at 
\begin{proposition} \label{prop-fay}
The shifted tau-functions satisfy 
\begin{eqnarray}
\label{FayT}
&&
    \mySum{\xi}{X'} 
    \myGamma{\xi}{\myset{Y''}}{\myset{X'}}
    \myalh{\tau}{\myset{X}/\xi}
    \myalh{\tau}{\xi\myset{Y}}
  + \mySum{\eta}{Y'} 
    \myGamma{\eta}{\myset{X''}}{\myset{Y'}}
    \myalh{\sigma}{\myset{X}\eta}
    \myalh{\rho}{\myset{Y}/\eta}
\nonumber\\ && \qquad
  = 
    \delta_{|\myset{X'}|,|\myset{Y''}|+1} 
    \myalh{\tau}{\myset{X}}
    \myalh{\tau}{\myset{Y}}
\hspace{36mm}
{\scriptstyle 
  (\mysetsize{X'}  \ge \mysetsize{Y''}+1, \; 
   \mysetsize{X''} \le \mysetsize{Y'}) 
},
\\[4mm]
\label{FayS}
&&
    \sum_{\xi \in \myset{X'}}     \myGamma{\xi}{\myset{Y''}}{\myset{X'}}
    \myalh{\tau}{\myset{X}/\xi}
    \myalh{\sigma}{\xi\myset{Y}}
  - \sum_{\eta \in \myset{Y'}}     \myGamma{\eta}{\myset{X''}}{\myset{Y'}}
    \myalh{\sigma}{\myset{X}\eta}
    \myalh{\tau}{\myset{Y}/\eta}
\nonumber\\ && \qquad
  = 
    \delta_{|\myset{X''}|,|\myset{Y'}|} 
    \myalh{\sigma}{\myset{X}}
    \myalh{\tau}{\myset{Y}}
  - \delta_{|\myset{X'}|,|\myset{Y''}|} 
    \myalh{\tau}{\myset{X}}
    \myalh{\sigma}{\myset{Y}}
\hspace{8mm}
{\scriptstyle 
  (\mysetsize{X'}  \ge \mysetsize{Y''}, \; 
   \mysetsize{X''} \le \mysetsize{Y'}) 
},
\\[4mm]
\label{FayR}
&&
    \sum_{\xi \in \myset{X'}}     \myGamma{\xi}{\myset{Y''}}{\myset{X'}}
    \myalh{\rho}{\myset{X}/\xi}
    \myalh{\tau}{\xi\myset{Y}}
  - \sum_{\eta \in \myset{Y'}}     \myGamma{\eta}{\myset{X''}}{\myset{Y'}}
    \myalh{\tau}{\myset{X}\eta}
    \myalh{\rho}{\myset{Y}/\eta}
\nonumber\\ && \qquad
  = 
    \delta_{|\myset{X''}|,|\myset{Y'}|} 
    \myalh{\tau}{\myset{X}}
    \myalh{\rho}{\myset{Y}}
  - \delta_{|\myset{X'}|,|\myset{Y''}|} 
    \myalh{\rho}{\myset{X}}
    \myalh{\tau}{\myset{Y}}
\hspace{8mm}
{\scriptstyle 
  (\mysetsize{X'}  \ge \mysetsize{Y''}, \; 
   \mysetsize{X''} \le |\myset{Y'}) 
}
\end{eqnarray}
where 
$\myset{X} = \myset{X'} \myset{X''}$ 
and 
$\myset{Y} = \myset{Y'} \myset{Y''}$. 
\end{proposition}

\noindent
From these general identities one can derive more simple ones, 
such as, for example, 
\begin{eqnarray}
&&
    \sum_{\xi \in \myset{X}}     
    \myGamma{\xi}{}{\myset{X}}
    \myalh{\tau}{\myset{X}/\xi}
    \myalh{\tau}{\xi\myset{Y}}
  + \sum_{\eta \in \myset{Y}}     
    \myGamma{\eta}{}{\myset{Y}}
    \myalh{\sigma}{\myset{X}\eta}
    \myalh{\rho}{\myset{Y}/\eta}
\nonumber\\ && \qquad
  = 
    \delta_{\mysetsize{X},1} \, 
    \myalh{\tau}{\myset{X}}
    \myalh{\tau}{\myset{Y}}
\hspace{58mm}
{\scriptstyle (\mysetsize{X} \ge 1,\; \mysetsize{Y} \ge 0)}
\label{FaySR}
\end{eqnarray}
where 
$\myGamma{\xi}{}{\myset{X}} = \myGamma{\xi}{\emptyset}{\myset{X}} 
= 1 / \Delta( \xi, \myset{X}/\xi)$, or 
\begin{equation}
\label{FayTT}
    \sum_{\xi \in \myset{X}}     
    \myGamma{\xi}{\myset{Y}}{\myset{X}}
    \myalh{\tau}{\myset{X}/\xi}
    \myalh{\tau}{\xi\myset{Y}}
  = 
    \delta_{\mysetsize{X},\mysetsize{Y}+1} \, 
    \myalh{\tau}{\myset{X}}
    \myalh{\tau}{\myset{Y}}
\hspace{19mm}
{\scriptstyle (\mysetsize{X} \ge \mysetsize{Y}+1)}, 
\end{equation}
\begin{equation}
\label{FayST}
    \sum_{\xi \in \myset{X}}     
    \myGamma{\xi}{\myset{Y}}{\myset{X}}
    \myalh{\tau}{\myset{X}/\xi}
    \myalh{\sigma}{\xi\myset{Y}}
  = 
    \myalh{\sigma}{\myset{X}}
    \myalh{\tau}{\myset{Y}}
  - \delta_{\mysetsize{X},\mysetsize{Y}} \, 
    \myalh{\tau}{\myset{X}}
    \myalh{\sigma}{\myset{Y}}
\hspace{5mm}
{\scriptstyle (\mysetsize{X} \ge \mysetsize{Y})},
\end{equation}
\begin{equation}
\label{FayTR}
    \sum_{\xi \in \myset{X}}     
    \myGamma{\xi}{\myset{Y}}{\myset{X}}
    \myalh{\rho}{\myset{X}/\xi}
    \myalh{\tau}{\xi\myset{Y}}
  = 
    \myalh{\tau}{\myset{X}}
    \myalh{\rho}{\myset{Y}}
  - \delta_{\mysetsize{X},\mysetsize{Y}} \, 
    \myalh{\rho}{\myset{X}}
    \myalh{\tau}{\myset{Y}}
\hspace{5mm} {\scriptstyle (\mysetsize{X} \ge \mysetsize{Y})}.
\end{equation}
Applying the simplest identities, which are equation \eref{FaySR} with 
$\myset{X}=\{\xi\}$ and $\myset{Y}=\{\eta\}$,
\begin{equation}
\label{FayBiT}
    \myalh{\tau}{}
    \myalh{\tau}{\xi\eta}
  + \myalh{\rho}{}
    \myalh{\sigma}{\xi\eta}
  = 
    \myalh{\tau}{\xi}
    \myalh{\tau}{\eta}
\end{equation}
and equations \eref{FayST} and \eref{FayTR} with 
$\myset{X}=\{\xi,\eta\}$ and $\myset{Y}=\emptyset$, 
\begin{equation}
\label{FayBiS}
    (\xi-\eta)
    \myalh{\tau}{}
    \myalh{\sigma}{\xi\eta}
  = 
    \myalh{\sigma}{\xi}
    \myalh{\tau}{\eta}
  - \myalh{\tau}{\xi}
    \myalh{\sigma}{\eta},
\end{equation}
\begin{equation}
\label{FayBiR}
    (\xi-\eta)
    \myalh{\rho}{}
    \myalh{\tau}{\xi\eta}
  = 
    \myalh{\tau}{\xi}
    \myalh{\rho}{\eta}
  - \myalh{\rho}{\xi}
    \myalh{\tau}{\eta}
\end{equation}
to \eref{FayTT}--\eref{FayTR} one can get a set of the `dual' ones, that 
cannot be obtained immediately from the general identities from proposition 
\ref{prop-fay}:
\begin{equation}
\label{FayTS}
    \sum_{\xi \in \myset{X}}     
    \myGamma{\xi}{\myset{Y}}{\myset{X}}
    \myalh{\sigma}{\myset{X}/\xi}
    \myalh{\tau}{\xi\myset{Y}}
  = 
  - \delta_{\mysetsize{X},\mysetsize{Y}+2} \, 
    \myalh{\sigma}{\myset{X}}
    \myalh{\tau}{\myset{Y}} 
\hspace{10mm}
{\scriptstyle (\mysetsize{X} \ge \mysetsize{Y}+2)}, 
\end{equation}
\begin{equation}
\label{FayRT}
    \sum_{\xi \in \myset{X}}     
    \myGamma{\xi}{\myset{Y}}{\myset{X}}
    \myalh{\tau}{\myset{X}/\xi}
    \myalh{\rho}{\xi\myset{Y}}
  = 
  - \delta_{\mysetsize{X},\mysetsize{Y}+2} \, 
    \myalh{\tau}{\myset{X}}
    \myalh{\rho}{\myset{Y}}
\hspace{10mm}
{\scriptstyle (\mysetsize{X} \ge \mysetsize{Y}+2)}, 
\end{equation}
\begin{equation}
\label{FayRS}
    \sum_{\xi \in \myset{X}} 
    \myGamma{\xi}{\myset{Y}}{\myset{X}}
    \myalh{\sigma}{\myset{X}/\xi}
    \myalh{\rho}{\xi\myset{Y}}
  = 
    \delta_{\mysetsize{X},\mysetsize{Y}+3} \, 
    \myalh{\sigma}{\myset{X}}
    \myalh{\rho}{\myset{Y}}
\hspace{10mm}
{\scriptstyle (\mysetsize{X} \ge \mysetsize{Y}+3)}. 
\end{equation}

Of course, the results presented in this section do not exhaust all possible 
identities that can be derived from \eref{innerTT}--\eref{defKRT}. 
However, they seem to be enough to obtain soliton solutions for almost all 
known integrable equations.

\section{Miwa shifts. \label{sec-miwa}}
%
Now we rewrite the identities of the previous section in terms of the so-called 
Miwa shifts. These shifts, which are defined for functions of an infinite 
number of variables, 
$f = f(\mathrm{t}) = f\left( t_{k} \right)_{k=1, ..., \infty}$ by 
\begin{equation}
  \myShift[E]\zeta f(\mathrm{t}) 
  = 
  f(\mathrm{t} + \epsilon [\zeta]) 
  = 
  f( t_{k} + \epsilon \zeta^{k}/k )_{k=1, ..., \infty} 
\end{equation}
where $\epsilon$ is a constant (usually, $1$ or $i$), seem to be one of the 
most natural tools to deal with integrable hierarchies. One can easily see 
from \eref{def-shits-AB} that $\myShift\xi$ cannot play the role of 
(or represent) the Miwa shifts because $\lim_{\xi \to 0}\myShift\xi \ne 1$. 
The simplest way to model the Miwa shifts, that was used in \cite{V14}, is to 
define them as 
\begin{equation}
  \myShift[E]{\alpha} 
  = 
  \myShift{\zeta(\alpha)} \myShift\kappa^{-1}, 
  \qquad
  \zeta(0)=\kappa 
\label{def-miwa}
\end{equation}
where $\kappa$ is some fixed parameter.
At first glance, $\kappa$ is not important because it can be included into 
the definition of the matrices $\mymatrix{L}$ and $\mymatrix{R}$ (that are 
still arbitrary). However, its role becomes crucial when one has to deal 
simultaneously with two (or, in principle, more) types of 
$\myShift[E]{}$-shifts (corresponding to different $\kappa$'s). 
A typical situation when this is the case is the problem of the so-called 
`negative' flows that has been studied in context of almost all integrable 
differential systems, when a hierarchy can be divided into two 
subhierarchies, which implies appearance of two families of differential 
operators and, hence, of two families of the Miwa shifts (each Miwa shift can 
be considered as an infinite series of differential operators). 
It turns out that these subhierarchies correspond to two 
possible choices of $\kappa$: 1) $\kappa$ is finite and 2) 
$\kappa \to \infty$. In \cite{V14}, we did not distinguish between these two 
cases. However, here, because of the different structure of the tau-functions, 
taking the $\kappa \to \infty$ limit needs more effort. So, we consider the 
$\myShift[E]{}$-shifts with $\kappa < \infty$ (that usually correspond to the 
`positive' or `classical' hierarchies) and with $\kappa=\infty$ 
(the `negative' flows) separately.

\subsection{Positive Miwa shifts. \label{sec-miwa-pos}}

As was mentioned previously, in the case of finite $\kappa$, this parameter 
can be eliminated by redefinition of the matrices $\mymatrix{L}$ and 
$\mymatrix{R}$. 
Thus, we define the positive Miwa shifts as 
\begin{equation}
  \myShifted[E]{\xi}{} = 
  \mathbb{T}_{\xi}\mathbb{T}_{0}^{-1} 
\label{def-miwa-pos}
\end{equation}
or, in terms of the soliton matrices, as 
\begin{equation}
  \begin{array}{lcl}
  \myShifted[E]{\xi}{\mymatrix{A}} 
  & = & 
  \mymatrix{A} 
  \left( \mymatrix{1} - \xi \mymatrix{R}^{-1} \right)^{-1}, 
  \\[2mm]
  \myShifted[E]{\xi}{\mymatrix{B}} 
  & = & 
  \mymatrix{B} 
  \left( \mymatrix{1} - \xi \mymatrix{L}^{-1} \right). 
  \end{array} 
\label{miwa-mat-pos}
\end{equation}
As in \cite{V14}, one can obtain a lot of Miwa-shift 
identities by sending part of $\myset{X}$ and $\myset{Y}$ 
in \eref{FayT}--\eref{FayR} to zero. 

For example, starting from equation \eref{FayTT} with 
$\mysetsize{X}=\mysetsize{Y}+2$ and applying 
$\myShift{0}^{-\mysetsize{X}+1}$, one arrives at 
\begin{equation}
  \mySum{\xi}{X} 
  \myGamma{\xi}{\myset{Y}}{\myset{X}} 
  \left( \myShifted[E]{\myset{X}/\xi}{\tau} \right)
  \left( \myShifted[E]{\xi\myset{Y}}{\tau}  \right)
  = 
  0
\hspace{36mm}
{\scriptstyle (\mysetsize{X} = \mysetsize{Y}+2)}
\end{equation}
which, after replacing 
$\myset{Y} \to \myset{Y} \cup \{ \overbrace{0..0}^{m} \}$, 
can be rewritten as 
\begin{equation}
  \mySum{\xi}{X} 
  \xi^{m}
  \myGamma{\xi}{\myset{Y}}{\myset{X}} 
  \left( \myShifted[E]{\myset{X}/\xi}{\tau} \right)
  \left( \myShifted[E]{\xi\myset{Y}}{\tau}  \right)
  = 
  0
\hspace{25mm} {\scriptstyle (\mysetsize{X} = \mysetsize{Y}+m+2)}. 
\end{equation}
At the same time, equation \eref{FayTT} with 
$\myset{X} \to \myset{X}\cup\{0\}$ and 
$\mysetsize{X}=\mysetsize{Y}+1$ leads to 
\begin{equation}
\fl\qquad
  \mySum{\xi}{X} 
  \xi^{-1}
  \myGamma{\xi}{\myset{Y}}{\myset{X}} 
  \left( \myShifted[E]{\myset{X}/\xi}{\tau} \right)
  \left( \myShifted[E]{\xi\myset{Y}}{\tau}  \right)
  = 
  (-)^{\mysetsize{X}-\mysetsize{Y}+1} 
  \frac{\pi(\myset{Y})}{\pi(\myset{X})} 
  \left( \myShifted[E]{\myset{X}}{\tau} \right)
  \left( \myShifted[E]{\myset{Y}}{\tau} \right) 
\hspace{3mm}{\scriptstyle (\mysetsize{X} = \mysetsize{Y}+1)}  
\end{equation}
where
\begin{equation}
  \pi(\myset{X}) 
  = 
  \prod_{\xi  \in \myset{X}} \xi. 
\label{def-pi}
\end{equation}
The preceding equations, as well as the ones that can be derived from 
\eref{FayST}, \eref{FayTR}, and \eref{FayTS}--\eref{FayRS} in a similar way, 
can be rewritten as follows:

\begin{proposition} \label{prop-miwa-pos}
For any sets $\myset{X}$ and $\myset{Y}$ restricted by  
\begin{equation}
  \mysetsize{X} \ge \mysetsize{Y}+1 
\end{equation}
the $\myShift[E]{}$-shifted tau-functions satisfy
\begin{equation}
\fl\qquad
  \mySum{\xi}{X} 
  \myHatGamma{\xi}{\myset{Y}}{\myset{X}} 
  \left( \myShifted[E]{\myset{X}/\xi}{\tau} \right)
  \left( \myShifted[E]{\xi\myset{Y}}{\tau}  \right)
  =  
  \gamma_{1}(\myset{Y},\myset{X}) 
  \left( \myShifted[E]{\myset{X}}{\tau} \right)
  \left( \myShifted[E]{\myset{Y}}{\tau} \right),
\end{equation}
\begin{equation}
\fl\qquad
  \mySum{\xi}{X} 
  \myHatGamma{\xi}{\myset{Y}}{\myset{X}} 
  \left( \myShifted[E]{\myset{X}/\xi}{\tau} \right)
  \left( \myShifted[E]{\xi\myset{Y}}{\sigma}  \right)
  = 
  \left( \myShifted[E]{\myset{X}}{\hat\sigma} \right)
  \left( \myShifted[E]{\myset{Y}}{\check\tau} \right)
  + 
  \gamma_{1}(\myset{Y},\myset{X}) 
  \left( \myShifted[E]{\myset{X}}{\tau}   \right)
  \left( \myShifted[E]{\myset{Y}}{\sigma} \right),
\end{equation}
\begin{equation}
\fl\qquad
  \mySum{\xi}{X} 
  \myHatGamma{\xi}{\myset{Y}}{\myset{X}} 
  \left( \myShifted[E]{\myset{X}/\xi}{\sigma} \right)
  \left( \myShifted[E]{\xi\myset{Y}}{\tau}    \right)
  = 
  - 
  \left( \myShifted[E]{\myset{X}}{\hat\sigma} \right)
  \left( \myShifted[E]{\myset{Y}}{\check\tau} \right)
  +
  \gamma_{1}(\myset{Y},\myset{X}) 
  \left( \myShifted[E]{\myset{X}}{\sigma} \right)
  \left( \myShifted[E]{\myset{Y}}{\tau}   \right), 
\end{equation}
\begin{equation}
\fl\qquad
  \mySum{\xi}{X} 
  \myHatGamma{\xi}{\myset{Y}}{\myset{X}} 
  \left( \myShifted[E]{\myset{X}/\xi}{\rho} \right)
  \left( \myShifted[E]{\xi\myset{Y}}{\tau}  \right)
  = 
  \left( \myShifted[E]{\myset{X}}{\hat\tau} \right)
  \left( \myShifted[E]{\myset{Y}}{\check\rho} \right)
  + 
  \gamma_{1}(\myset{Y},\myset{X}) 
  \left( \myShifted[E]{\myset{X}}{\rho}   \right)
  \left( \myShifted[E]{\myset{Y}}{\tau} \right), 
\end{equation}
\begin{equation}
\fl\qquad
  \mySum{\xi}{X} 
  \myHatGamma{\xi}{\myset{Y}}{\myset{X}} 
  \left( \myShifted[E]{\myset{X}/\xi}{\tau} \right)
  \left( \myShifted[E]{\xi\myset{Y}}{\rho}    \right)
  = 
  - 
  \left( \myShifted[E]{\myset{X}}{\hat\tau} \right)
  \left( \myShifted[E]{\myset{Y}}{\check\rho} \right)
  +
  \gamma_{1}(\myset{Y},\myset{X}) 
  \left( \myShifted[E]{\myset{X}}{\tau} \right)
  \left( \myShifted[E]{\myset{Y}}{\rho}   \right)
\end{equation}
where we use the designations 
\begin{equation}
  \myHatGamma{\xi}{\myset{Y}}{\myset{X}} 
  = 
  \xi^{ \mysetsize{X} - \mysetsize{Y} - 2}
  \myGamma{\xi}{\myset{Y}}{\myset{X}},
\label{def-hat-Gamma}
\end{equation}
\begin{equation}
  \gamma_{n}(\myset{Y},\myset{X}) 
  = 
  (-)^{n-1}
  \delta_{\mysetsize{X},\mysetsize{Y}+n} \,
  \frac{ \pi(\myset{Y}) }{ \pi(\myset{X}) }
\label{def-gamma}
\end{equation}
and 
\begin{equation}
  \hat\omega = \mathbb{T}_{0} \, \omega, 
  \qquad
  \check\omega = \mathbb{T}_{0}^{-1} \omega. 
\end{equation}
\end{proposition}

\subsection{Negative Miwa shifts. \label{sec-miwa-neg}}

The negative Miwa shifts, that correspond to $\kappa \to \infty$ and that will 
be denoted by the symbol $\myBarShift[E]{\xi}$, are easy to define in terms 
of the soliton matrices, 
\begin{equation}
  \begin{array}{lcl}
  \myBarShifted[E]{\xi}{\mymatrix{A}} 
  & = & 
  \mymatrix{A} 
  \left( \mymatrix{1} - \xi \mymatrix{R} \right)^{-1}, 
  \\[2mm]
  \myBarShifted[E]{\xi}{\mymatrix{B}} 
  & = & 
  \mymatrix{B} 
  \left( \mymatrix{1} - \xi \mymatrix{L} \right). 
  \end{array} 
\label{miwa-mat-neg}
\end{equation}
However, to be able to use the $\myShift{}$-formulae from section 
\ref{sec-fay}, one has to rewrite this definition separately for each 
tau-function as 
\begin{equation}
  \begin{array}{lcl} 
	\myBarShifted[E]{\xi}{\tau} &=& \myShifted{1/\xi}{\,\tau}, 
  \\
	\myBarShifted[E]{\xi}{\sigma} &=& - \xi^{-1} \; \myShifted{1/\xi}{\,\sigma}, 
  \\
	\myBarShifted[E]{\xi}{\rho} &=& - \xi \; \myShifted{1/\xi}{\,\rho}
  \end{array}
\label{def-miwa-neg}
\end{equation}
or 
\begin{equation}
  \begin{array}{lcl} 
	\myBarShifted[E]{\myset{X}}{\tau} 
	& = & 
	\myShifted{\widetilde{\myset{X}}}{\,\tau}, 
  \\[2mm]
	\myBarShifted[E]{\myset{X}}{\sigma} 
	& = & \displaystyle 
	(-)^{\mysetsize{X}} \frac{ 1 }{ \pi(\myset{X}) } \; 
	\myShifted{\widetilde{\myset{X}}}{\,\sigma}, 
  \\[4mm]
	\myBarShifted[E]{\myset{X}}{\rho} 
	& = & 
  (-)^{\mysetsize{X}} \pi(\myset{X}) \; 
  \myShifted{\widetilde{\myset{X}}}{\,\rho}
  \end{array}
\end{equation}
where 
\begin{equation}
  \widetilde{\myset{X}} 
  = 
  \bigcup_{\xi\in\myset{X}} \xi^{-1} 
\end{equation}
and $\pi(\myset{X})$ is given by \eref{def-pi}. 

The identities presented next can be obtained from equations 
\eref{FayTT}--\eref{FayTR} and \eref{FayTS}--\eref{FayRS} with 
$\myset{X}$ and $\myset{Y}$ being replaced by $\widetilde{\myset{X}}$ and 
$\widetilde{\myset{Y}}$ by calculations similar to the ones that we 
carried out in the case of the positive Miwa shifts. As a result, one can 
obtain 
\begin{proposition} \label{prop-miwa-neg}
The $\myBarShifted[E]{}{}$-shifted tau-functions satisfy
\begin{equation}
\fl
  \mySum{\xi}{X} 
  \myHatGamma{\xi}{\myset{Y}}{\myset{X}} 
  \left( \myBarShifted[E]{\myset{X}/\xi}{\tau} \right)
  \left( \myBarShifted[E]{\xi\myset{Y}}{\tau}  \right)
  = 
  \gamma_{1}(\myset{Y},\myset{X}) \; 
  \left( \myBarShifted[E]{\myset{X}}{\tau} \right)
  \left( \myBarShifted[E]{\myset{Y}}{\tau} \right)
\hfill
{\scriptstyle (\mysetsize{X} \ge \mysetsize{Y}+1)}, 
\hspace{2mm}
\end{equation}
\begin{equation}
\fl
  \mySum{\xi}{X} 
  \xi  
  \myHatGamma{\xi}{\myset{Y}}{\myset{X}} 
  \left( \myBarShifted[E]{\myset{X}/\xi}{\tau}  \right)
  \left( \myBarShifted[E]{\xi\myset{Y}}{\sigma} \right)
  = 
  \left( \myBarShifted[E]{\myset{X}}{\sigma} \right)
  \left( \myBarShifted[E]{\myset{Y}}{\tau}  \right)
  +
  \gamma_{0}(\myset{Y},\myset{X}) \; 
  \left( \myBarShifted[E]{\myset{X}}{\tau}   \right)
  \left( \myBarShifted[E]{\myset{Y}}{\sigma} \right)
\hfill
{\scriptstyle (\mysetsize{X} \ge \mysetsize{Y})}, 
\hspace{2mm}
\end{equation}
\begin{equation}
\fl
  \mySum{\xi}{X} 
  \xi^{-1} 
  \myHatGamma{\xi}{\myset{Y}}{\myset{X}} 
  \left( \myBarShifted[E]{\myset{X}/\xi}{\sigma} \right)
  \left( \myBarShifted[E]{\xi\myset{Y}}{\tau}    \right)
  = 
  \gamma_{2}(\myset{Y},\myset{X}) \; 
	\frac{ \pi(\myset{Y}) }{ \pi(\myset{X}) } \; 
  \left( \myBarShifted[E]{\myset{X}}{\sigma} \right)
  \left( \myBarShifted[E]{\myset{Y}}{\tau}   \right)
\hfill
{\scriptstyle (\mysetsize{X} \ge \mysetsize{Y}+2)}, 
\hspace{2mm}
\end{equation}
\begin{equation}
\fl
  \mySum{\xi}{X} 
  \xi  
  \myHatGamma{\xi}{\myset{Y}}{\myset{X}} 
  \left( \myBarShifted[E]{\myset{X}/\xi}{\rho} \right)
  \left( \myBarShifted[E]{\xi\myset{Y}}{\tau}  \right)
  = 
  \left( \myBarShifted[E]{\myset{X}}{\tau} \right)
  \left( \myBarShifted[E]{\myset{Y}}{\rho} \right)
  + 
  \gamma_{0}(\myset{Y},\myset{X}) \; 
  \left( \myBarShifted[E]{\myset{X}}{\rho} \right)
  \left( \myBarShifted[E]{\myset{Y}}{\tau} \right)
\hfill
{\scriptstyle (\mysetsize{X} \ge \mysetsize{Y})}, 
\hspace{2mm}
\end{equation}
\begin{equation}
\fl
  \mySum{\xi}{X} 
  \xi^{-1} 
  \myHatGamma{\xi}{\myset{Y}}{\myset{X}} 
  \left( \myBarShifted[E]{\myset{X}/\xi}{\tau} \right)
  \left( \myBarShifted[E]{\xi\myset{Y}}{\rho}  \right)
  = 
  \gamma_{2}(\myset{Y},\myset{X}) \; 
	\frac{ \pi(\myset{Y}) }{ \pi(\myset{X}) } \; 
  \left( \myBarShifted[E]{\myset{X}}{\tau} \right)
  \left( \myBarShifted[E]{\myset{Y}}{\rho} \right)
\hfill
{\scriptstyle (\mysetsize{X} \ge \mysetsize{Y}+2)} 
\hspace{3mm}
\end{equation}
where $\myHatGamma{\xi}{\myset{Y}}{\myset{X}}$ and 
$\gamma_{n}(\myset{Y},\myset{X})$ are defined in 
\eref{def-hat-Gamma} and \eref{def-gamma}.
\end{proposition}

\section{Differential Fay identities. \label{sec-diff}}
%
In this section we derive the differential consequences of the general Fay 
identities presented in section \ref{sec-fay}. As in the case of the Miwa 
shifts, we consider the positive and negative cases separately, because 
of some differences in the procedure of taking the corresponding limits.

\subsection{Positive differential identities. }

Following \cite{V14}, we define the differential operator $\partial_{\lambda}$ 
by 
\begin{equation}
  \partial_{\lambda} 
  = 
  \lim_{\xi\to\lambda} 
  \frac{ 1 }{ \xi - \lambda } 
  \left( \myShift[E]{\xi} \myShift[E]{\lambda}^{-1} - 1 \right) 
  = 
  \lim_{\xi\to\lambda} 
  \frac{ 1 }{ \xi - \lambda } 
  \left( \myShift{\xi} \myShift{\lambda}^{-1} - 1 \right). 
\label{partial-def}
\end{equation}
Now, we do not repeat the calculations of \cite{V14} but take a little bit 
different approach that leads to more transparent (and shorter) formulae. 
The idea is to use the general identities from proposition \ref{prop-fay} 
(which are absent in \cite{V14}) and take one of the summation sets, 
$\myset{X'}$ or $\myset{Y'}$, consisting of only two elements, say 
$\{\gamma,\lambda\}$. Then, applying the shift $\myShift{\lambda}^{-1}$ and 
sending $\gamma$ to $\lambda$ we obtain the desired result.
For example, starting from \eref{FayT} with 
$\myset{X'}=\{\gamma,\lambda\}$ and $\myset{Y''}=\emptyset$ 
(note that the size of the set $\myset{Y''}$ is bounded by the size of the 
set $\myset{X'}$), the first sum becomes (after $\myShift{\lambda}^{-1}$ 
shift)
\begin{equation}
  \frac{ 1 }{ \gamma -\lambda } 
  \left( 
    \myalh{\tau}{\myset{X''}}
    \myalh{\tau}{\gamma \bar{\lambda} \myset{Y'}}
    - 
    \myalh{\tau}{\gamma \bar{\lambda} \myset{X''}}
    \myalh{\tau}{\myset{Y'}} 
  \right) 
\end{equation}
which in the $\gamma\to\lambda$ limit is 
$-\myHirota{\lambda}{\myalh{\tau}{\myset{X''}}}{\myalh{\tau}{\myset{Y'}}}$ 
where $D_{\lambda}$ stands for the Hirota bilinear operator  
\begin{equation}
  \myHirota{\lambda}{u}{v} 
  = 
  \left( \partial_{\lambda} u \right) v 
  - 
  u \left( \partial_{\lambda} v \right). 
\end{equation}
Thus, \eref{FayT} becomes the expression for  
$\myHirota{\lambda}{\myalh{\tau}{\myset{X}}}{\myalh{\tau}{\myset{Y}}}$ 
in terms of a sum of bilinear combinations of $\sigma$ and $\rho$.
Repeating this procedure with \eref{FayS} and \eref{FayR} one can obtain the 
following result. 

\begin{proposition} \label{prop-diff-fay-pos}
For any sets $\myset{X}$ and $\myset{Y}$ restricted by  
\begin{equation}
  \mysetsize{X} \ge \mysetsize{Y} 
\end{equation}
the bilinear derivatives of the shifted tau-functions can be presented as 
\begin{eqnarray}
  \myHirota{\lambda}{\myalh{\tau}{\myset{X}}}{\myalh{\tau}{\myset{Y}}} 
  & = & 
  - \sum_{\xi \in \myset{X}} 
    \myGamma{\xi}{\myset{Y}}{\myset{X}}
    \myalh{\rho}{\bar{\lambda}\myset{X}/\xi}
    \myalh{\sigma}{\lambda\xi\myset{Y}}, 
\\
  \myHirota{\lambda}{\myalh{\sigma}{\myset{X}}}{\myalh{\tau}{\myset{Y}}}
  & = & 
  \sum_{\xi \in \myset{X}} 
    \myGamma{\xi}{\myset{Y}}{\myset{X}}
    \myalh{\tau}{\bar{\lambda}\myset{X}/\xi} 
    \myalh{\sigma}{\lambda\xi\myset{Y}}
  + 
  \delta_{\mysetsize{X},\mysetsize{Y}} 
    \myalh{\tau}{\bar{\lambda}\myset{X}} 
    \myalh{\sigma}{\lambda\myset{Y}}, 
\\
  \myHirota{\lambda}{\myalh{\tau}{\myset{X}}}{\myalh{\rho}{\myset{Y}}}
  & = & 
  \sum_{\xi \in \myset{X}} 
    \myGamma{\xi}{\myset{Y}}{\myset{X}}
    \myalh{\rho}{\bar{\lambda}\myset{X}/\xi} 
    \myalh{\tau}{\lambda\xi\myset{Y}}
  + 
  \delta_{\mysetsize{X},\mysetsize{Y}} 
    \myalh{\rho}{\bar{\lambda}\myset{X}} 
    \myalh{\tau}{\lambda\myset{Y}}. 
\end{eqnarray}
\end{proposition}

The simplest differential identities can be written as
\begin{equation}
  \myHirota{\lambda}{\myalh{\tau}{\xi}}{\myalh{\tau}{}}
  = 
  - 
  \myalh{\sigma}{\lambda\xi}
  \myalh{\rho}{\bar{\lambda}}
\end{equation}
and
\begin{equation}
  \begin{array}{lcl}
  \myHirota{\lambda}{\myalh{\sigma}{}}{\myalh{\tau}{}}
  & = & 
  \myalh{\sigma}{\lambda}
  \myalh{\tau}{\bar{\lambda}}, 
  \\
  \myHirota{\lambda}{\myalh{\tau}{}}{\myalh{\rho}{}}
  & = & 
  \myalh{\tau}{\lambda}
  \myalh{\rho}{\bar{\lambda}}. 
  \end{array} 
\label{diff-pos-str}
\end{equation}

\subsection{Negative differential identities. }

In this subsection we discuss the derivatives of the tau-functions with respect 
to the negative differential operator 
\begin{equation}
  \bar\partial_{\mu} 
  = 
  \lim_{\xi\to\mu} 
  \frac{ 1 }{ \xi - \mu } 
  \left( \myBarShift[E]{\xi} \myBarShift[E]{\mu}^{-1} - 1 \right).  
\end{equation}
The calculations we need to make are similar to the ones for the case of the 
positive flows. However, we have to take into account that shift 
$\myBarShift[E]{\mu}$ corresponds to $\myShift{1/\mu}$. 
To make the following formulae more readable, we use throughout this section 
the designation $\lambda = 1/\mu$, rewrite the definition of 
$\bar\partial_{\mu}$ operator as 
\begin{equation}
  \bar\partial_{\mu} 
  = 
  - \lambda^{2} 
  \lim_{\gamma\to\lambda} 
  \frac{ 1 }{ \gamma - \lambda } 
  \left( \myShift{\gamma} \myShift{\lambda}^{-1} - 1 \right), 
  \qquad
  \lambda=1/\mu  
\end{equation}
and proceed as in the previous subsection: we start from the identities 
\eref{FayT}--\eref{FayR} with two-element set 
$\myset{X'}$, $\myset{X'}=\{\gamma,\lambda\}$, apply 
$\myShift{\lambda}^{-1}$ and 
send $\gamma$ to $\lambda$ using the expansions 
\begin{eqnarray}
  \myalh{\tau}{\gamma\bar{\lambda}}
  & = & 
    \tau 
  - \mu^{2}
    (\gamma-\lambda)
    \bar\partial_{\mu} \tau 
  + o(\gamma-\lambda), 
\\[2mm] 
  \myalh{\sigma}{\gamma\bar{\lambda}}
  & = & 
    \sigma 
  - 
  (\gamma-\lambda) \left[
    \mu^{2}
    \bar\partial_{\mu} \sigma 
    + \mu \sigma 
  \right]
  + o(\gamma-\lambda), 
\\[2mm] 
  \myalh{\rho}{\gamma\bar{\lambda}}
  & = & 
  \rho 
  - 
  (\gamma-\lambda) \left[ 
    \mu^{2}
    \bar\partial_{\mu} \rho 
    - \mu \rho 
  \right] 
  + o(\gamma-\lambda). 
\end{eqnarray}
As a result, one can obtain for $\mysetsize{X} \ge \mysetsize{Y}$
\begin{eqnarray}
  \mu^{2} \, 
  \myBarHirota{\mu}{\myalh{\tau}{\myset{X}}}{\myalh{\tau}{\myset{Y}}} 
  & = & 
  \sum_{\xi \in \myset{X}} 
    \myGamma{\xi}{\myset{Y}}{\myset{X}}
    \myalh{\rho}{\bar{\lambda}\myset{X}/\xi} 
    \myalh{\sigma}{\lambda\xi\myset{Y}},
\label{diff-neg-tt}
\\
  \mu^{2} \, 
  \myBarHirota{\mu}{\myalh{\sigma}{\myset{X}}}{\myalh{\tau}{\myset{Y}}} 
  & = & 
  - \sum_{\xi \in \myset{X}} 
      \myGamma{\xi}{\myset{Y}}{\myset{X}}
      \myalh{\tau}{\bar{\lambda}\myset{X}/\xi}
      \myalh{\sigma}{\lambda\xi\myset{Y}}
\nonumber \\&&
  - \delta_{\mysetsize{X},\mysetsize{Y}} 
    \myalh{\tau}{\bar{\lambda}\myset{X}}
    \myalh{\sigma}{\lambda\myset{Y}}
  - \mu 
    \myalh{\sigma}{\myset{X}}
    \myalh{\tau}{\myset{Y}}, 
\label{diff-neg-st}
\\[2mm]
  \mu^{2} \, 
  \myBarHirota{\mu}{\myalh{\tau}{\myset{X}}}{\myalh{\rho}{\myset{Y}}} 
  & = & 
  - \sum_{\xi \in \myset{X}} 
      \myGamma{\xi}{\myset{Y}}{\myset{X}}
      \myalh{\rho}{\bar{\lambda}\myset{X}/\xi}
      \myalh{\tau}{\lambda\xi\myset{Y}}
\nonumber \\&&
  - \delta_{\mysetsize{X},\mysetsize{Y}} 
    \myalh{\rho}{\bar{\lambda}\myset{X}}
    \myalh{\tau}{\lambda\myset{Y}}
  - \mu
    \myalh{\tau}{\myset{X}}
    \myalh{\rho}{\myset{Y}}
\label{diff-neg-tr}
\end{eqnarray}
where 
$ \myBarHirota{\mu}{u}{v} 
  = 
  \left( \bar\partial_{\mu} u \right) v 
  - 
  u \left( \bar\partial_{\mu} v \right). 
$
These equations are the negative analogues of the ones collected in 
proposition \ref{prop-diff-fay-pos}. However, they possess a serious drawback: 
they do not have clear $\mu \to 0$ limit. 
In terms of functions of an infinite number of variables (or, if we consider 
both positive and negative flows simultaneously, functions of two infinite 
sets of variables, 
$f=f( t_{j}, \bar{t}_{k} )_{j,k=1, ..., \infty}$) 
this means that to see the action of a particular differential operator 
$\partial/\partial\bar{t}_{k}$ one needs some additional work 
(even for the simplest one, $\partial/\partial\bar{t}_{1}$). 
It turns out, that this problem can be solved 
by replacing $\myShift{\lambda}^{\pm 1}$ 
in the right-hand sides of \eref{diff-neg-tt}-\eref{diff-neg-tr} 
with $\myShift[E]{\mu}^{\pm 1}$, which leads to the following result 
(we omit the details of these calculations). 

\begin{proposition} \label{prop-diff-fay-neg}
For any sets $\myset{X}$ and $\myset{Y}$ 
the bilinear derivatives of the shifted tau-functions with respect to the 
negative variables can be presented as 
\begin{eqnarray}
  \myBarHirota{\mu}{\myalh{\tau}{\myset{X}}}{\myalh{\tau}{\myset{Y}}} 
  & = & 
  \sum_{\xi \in \myset{X}} 
    \myGamma{\xi}{\myset{Y}}{\myset{X}}
    \left( \myBarShiftInv[E]{\mu} \myalh{\rho}{\myset{X}/\xi}  \right) 
    \left( \myBarShift[E]{\mu}    \myalh{\sigma}{\xi\myset{Y}} \right)
\hspace{6mm}
{\scriptstyle (\mysetsize{X} \ge \mysetsize{Y})}, 
\\[2mm]
  \myBarHirota{\mu}{\myalh{\sigma}{\myset{X}}}{\myalh{\tau}{\myset{Y}}} 
  & = & 
  \sum_{\xi \in \myset{X}} 
    \xi \myGamma{\xi}{\myset{Y}}{\myset{X}}
    \left( \myBarShiftInv[E]{\mu} \myalh{\tau}{\myset{X}/\xi}  \right)
    \left( \myBarShift[E]{\mu}    \myalh{\sigma}{\xi\myset{Y}} \right) 
\nonumber\\&&
  + \delta_{\mysetsize{X},\mysetsize{Y}+1} 
    \left( \myBarShiftInv[E]{\mu} \myalh{\tau}{\myset{X}}   \right) 
    \left( \myBarShift[E]{\mu}    \myalh{\sigma}{\myset{Y}} \right) 
\hspace{12mm}
{\scriptstyle (\mysetsize{X} \ge \mysetsize{Y}+1)}, 
\\[2mm] 
  \myBarHirota{\mu}{\myalh{\tau}{\myset{X}}}{\myalh{\rho}{\myset{Y}}} 
  & = & 
  \sum_{\xi \in \myset{X}} 
    \xi \myGamma{\xi}{\myset{Y}}{\myset{X}}
    \left( \myBarShiftInv[E]{\mu} \myalh{\rho}{\myset{X}/\xi} \right)
    \left( \myBarShift[E]{\mu}    \myalh{\tau}{\xi\myset{Y}}  \right) 
\nonumber\\&&
  + \delta_{\mysetsize{X},\mysetsize{Y}+1} 
    \left( \myBarShiftInv[E]{\mu} \myalh{\rho}{\myset{X}} \right) 
    \left( \myBarShift[E]{\mu}    \myalh{\tau}{\myset{Y}} \right) 
\hspace{12mm}
{\scriptstyle (\mysetsize{X} \ge \mysetsize{Y}+1)}. 
\end{eqnarray}
\end{proposition}

It is easy to see that from the preceding identities one cannot obtain simple 
examples similar to \eref{diff-pos-str}. This is a manifestation of the 
fact that the negative flows are usually non-local. There are very few known 
integrable systems with local negative equations. One of them is the 
AL model \cite{AL75,AL76,AS,APT}, which we discuss in the next section.

\section{Soliton solutions for the ALH. \label{sec-app}}

As an example of an application of the Fay identities discussed in this paper, 
we derive in this section the $N$-bright soliton solutions for the ALH. 
The ALH is an infinite set of ordinal differential-difference equations, that 
has been introduced by Ablowitz and Ladik in 1975 \cite{AL75,AL76,AS,APT}. 
The most well-known of these equations is the discrete nonlinear Schr\"odinger 
equation.
All equations of the ALH can be presented as 
\begin{equation}
  \begin{array}{lcl}
  \partial q_{n} / \partial t_{j} 
  & = & 
  Q^{(j)}\left( q_{n}, r_{n}, q_{n \pm 1}, r_{n \pm 1}, ... \right), 
	\\
	\partial r_{n} / \partial t_{j} 
  & = & 
  R^{(j)}\left( q_{n}, r_{n}, q_{n \pm 1}, r_{n \pm 1}, ... \right)
	\end{array}
\end{equation}
where $Q^{(j)}$ and $R^{(j)}$ are some polynomials whose form (and order) 
depends on which equation of the hierarchy one deals with. 
In what follows, we use the so-called functional representation of the ALH, 
which gives us a possibility to derive solutions for all  equations of the 
hierarchy simultaneously. This approach consists of considering functions 
$q_{n}$ and $r_{n}$ as functions of two infinite number of variables 
(that describe `positive' and `negative' subhierarchies),
\begin{equation}
  \begin{array}{lclcl}
  q_{n} 
  & = & 
  q_{n}\left( \mathrm{z}, \mathrm{\bar{z}} \right) 
  & = & 
  q\left( z_{j}, \bar{z}_{k} \right)_{j,k=1, ..., \infty}, 
	\\
  r_{n} 
  & = & 
  r_{n}\left( \mathrm{z}, \mathrm{\bar{z}} \right) 
  & = & 
  r\left( z_{j}, \bar{z}_{k} \right)_{j,k=1, ..., \infty}, 
	\end{array}
\end{equation}
and to rewrite the ALH using the Miwa shifts, 
$\myShift[E]{\xi}$ and $\myBarShift[E]{\eta}$, that are defined as 
\begin{equation}
  \begin{array}{l}
  \myShift[E]\xi \omega_{n} =  
  \omega_{n}\left( 
    z_{j} + i \xi^{j}/j, \bar{z}_{k} 
  \right)_{j,k=1, ..., \infty}, 
  \\[2mm] 
  \myBarShift[E]\eta \omega_{n} =  
  \omega_{n}\left( 
    z_{j}, \bar{z}_{k}  + i \eta^{k}/k 
  \right)_{j,k=1, ..., \infty}, 
  \end{array} 
\label{alh-miwa}
\end{equation}
as a finite system of functional-difference equations
\begin{eqnarray}
\label{alh-fr-pos-q} 
  ( \myShift[E]{\xi} - 1 ) q_{n} 
  & = & 
  \xi 
  \left[ 1 - (\myShifted[E]{\xi}{q_{n}})r_{n} \right] 
  \myShifted[E]{\xi}{q_{n+1}}, 
\\
  ( 1 - \myShift[E]{\xi} ) r_{n} 
  & = & 
  \xi 
  \left[ 1 - (\myShifted[E]{\xi}{q_{n}})r_{n} \right] 
  r_{n-1} 
\end{eqnarray}
(the positive subhierarchy) and 
\begin{eqnarray}
  ( \myBarShift[E]{\eta} - 1 ) q_{n} 
  & = & 
  \eta 
  \left[ 1 - (\myBarShifted[E]{\eta}{q_{n}})r_{n} \right] 
  \myBarShifted[E]{\eta}{q_{n-1}}, 
\\
  ( 1 - \myBarShift[E]{\eta} ) r_{n} 
  & = & 
  \eta 
  \left[ 1 - (\myBarShifted[E]{\eta}{q_{n}})r_{n} \right] 
  r_{n+1} 
\label{alh-fr-neg-r} 
\end{eqnarray}
(the negative one) from which one can obtain `traditional' 
differential-difference equations by the power series expansion in $\xi$ 
and $\eta$.
A reader can find detailed derivation of the functional representation of the ALH 
in \cite{V98,V02}. 

Introducing the tau-functions $\tau_{n}$, $\sigma_{n}$ and $\rho_{n}$ by
\begin{equation}
  q_{n} = \frac{\sigma_{n}}{\tau_{n}}, 
  \qquad
  r_{n} = \frac{\rho_{n}}{\tau_{n}} 
\end{equation}
together with the restriction 
\begin{equation}
  \tau_{n-1}
  \tau_{n+1} 
  = 
  \tau_{n}^{2} 
  - 
  \rho_{n}
  \sigma_{n} 
\label{alh-main}
\end{equation}
one arrives at the bilinear system 
\begin{eqnarray}
  \tau_{n-1}
  \myShifted[E]{\xi}{\tau_{n+1}} 
  - 
  \tau_{n}
  \myShifted[E]{\xi}{\tau_{n}} 
  + 
  \rho_{n}
  \myShifted[E]{\xi}{\sigma_{n}} 
  = 0, 
\label{alh-pos-a}
\\
  \xi 
  \tau_{n-1}
  \myShifted[E]{\xi}{\sigma_{n+1}} 
  - 
  \tau_{n}
  \myShifted[E]{\xi}{\sigma_{n}} 
  + 
  \sigma_{n}
  \myShifted[E]{\xi}{\tau_{n}} 
  = 0, 
\label{alh-pos-b}
\\
  \xi 
  \rho_{n-1}
  \myShifted[E]{\xi}{\tau_{n+1}} 
  - 
  \rho_{n}
  \myShifted[E]{\xi}{\tau_{n}} 
  + 
  \tau_{n}
  \myShifted[E]{\xi}{\rho_{n}} 
  = 0 
\label{alh-pos-c}
\end{eqnarray}
(the positive subhierarchy) and 
\begin{eqnarray}
  \tau_{n}
  \myBarShifted[E]{\eta}{\tau_{n}} 
  - 
  \tau_{n+1}
  \myBarShifted[E]{\eta}{\tau_{n-1}} 
  - 
  \rho_{n}
  \myBarShifted[E]{\eta}{\sigma_{n}} 
  = 0, 
\label{alh-neg-a}
\\
  \eta
  \tau_{n+1}
  \myBarShifted[E]{\eta}{\sigma_{n-1}} 
  - 
  \tau_{n}
  \myBarShifted[E]{\eta}{\sigma_{n}} 
  + 
  \sigma_{n}
  \myBarShifted[E]{\eta}{\tau_{n}} 
  = 0, 
\label{alh-neg-b}
\\
  \eta
  \rho_{n+1}
  \myBarShifted[E]{\eta}{\tau_{n-1}} 
  - 
  \rho_{n}
  \myBarShifted[E]{\eta}{\tau_{n}} 
  + 
  \tau_{n}
  \myBarShifted[E]{\eta}{\rho_{n}} 
  = 0 
\label{alh-neg-c}
\end{eqnarray}
(the negative one).
Namely these equations we solve in this section by means of the soliton Fay 
identities. To solve \eref{alh-pos-a}--\eref{alh-neg-c}  
we need the following ones: 
equations \eref{FayS} and \eref{FayT} with 
$\myset{X}' = \{\zeta,\kappa\}$, $\myset{X}'' = \emptyset$, 
$\myset{Y}' = \{\kappa\}$ and $\myset{Y}'' = \emptyset$, 
\begin{eqnarray}
    \left( \zeta - \kappa \right)
    \myalh{\tau}{\bar{\kappa}}
    \myalh{\sigma}{\kappa\zeta}
  - \myalh{\tau}{}
    \myalh{\sigma}{\zeta}
  + \myalh{\sigma}{\kappa}
    \myalh{\tau}{\bar{\kappa}\zeta}
  = 
  0,
\label{alh-fay-a}
\\
  \left( \zeta - \kappa \right) 
  \myalh{\rho}{\bar{\kappa}}
  \myalh{\sigma}{\kappa\zeta}
  - 
  \myalh{\tau}{\kappa}
  \myalh{\tau}{\bar{\kappa}\zeta}
  + 
  \myalh{\tau}{}
  \myalh{\tau}{\zeta}
  = 
  0,
\label{alh-fay-b}
\end{eqnarray}
equations \eref{FayBiS}, \eref{FayBiT} with $\xi=\zeta$ and $\eta=\kappa$,
\begin{eqnarray}
  \left( \zeta - \kappa \right) 
  \myalh{\tau}{}
  \myalh{\sigma}{\kappa\zeta}
  - 
  \myalh{\tau}{\kappa}
  \myalh{\sigma}{\zeta}
  + 
  \myalh{\sigma}{\kappa}
  \myalh{\tau}{\zeta}
  = 
  0, 
\label{alh-fay-c}
\\
  \myalh{\tau}{}
  \myalh{\tau}{\kappa\zeta}
  + 
  \myalh{\rho}{}
  \myalh{\sigma}{\kappa\zeta}
  - 
  \myalh{\tau}{\kappa}
  \myalh{\tau}{\zeta}
  = 
  0
\label{alh-fay-d}
\end{eqnarray}
and \eref{FayBiT} with $\xi=\eta=\kappa$, 
\begin{equation}
  0
  = 
  \tau^{2}
  - 
  \myalh{\tau}{\kappa}
  \myalh{\tau}{\bar{\kappa}}
  - 
  \myalh{\rho}{\bar{\kappa}}
  \myalh{\sigma}{\kappa}.
\end{equation}
Comparing the last equation with \eref{alh-main}, it is easy to conclude 
that the natural way to introduce the $n$-dependence is 
\begin{equation}
  \tau_{n}
  = 
  \myShift{\kappa}^{n} \tau,
\qquad
  \sigma_{n}
  = 
  \myShift{\kappa}^{n+1} \sigma,
\qquad
  \rho_{n}
  = 
  \myShift{\kappa}^{n-1} \rho, 
\label{alh-n}
\end{equation}
which transforms the remaining equations, \eref{alh-fay-a}--\eref{alh-fay-d}, 
into 
\begin{eqnarray}
  \left( \zeta - \kappa \right)
  \tau_{n}
  \myalh{\sigma_{n+1}}{\zeta}
  - 
  \tau_{n+1}
  \myalh{\sigma_{n}}{\zeta}
  + 
  \sigma_{n+1}
  \myalh{\tau_{n}}{\zeta}
  = 0, 
\label{alh-nz-a}
\\
  \left( \zeta - \kappa \right)
  \rho_{n}
  \myalh{\sigma_{n}}{\zeta}
  - 
  \tau_{n+1}
  \myalh{\tau_{n-1}}{\zeta}
  + 
  \tau_{n}
  \myalh{\tau_{n}}{\zeta}
  = 0, 
\label{alh-nz-b}
\\
  \left( \zeta - \kappa \right)
  \tau_{n}
  \myalh{\sigma_{n}}{\zeta}
  - 
  \tau_{n+1}
  \myalh{\sigma_{n-1}}{\zeta}
  + 
  \sigma_{n}
  \myalh{\tau_{n}}{\zeta}
  = 0, 
\label{alh-nz-c}
\\
  \tau_{n}
  \myalh{\tau_{n+1}}{\zeta}
  - 
  \tau_{n+1}
  \myalh{\tau_{n}}{\zeta}
  + 
  \rho_{n+1}
  \myalh{\sigma_{n}}{\zeta}
  = 0. 
\label{alh-nz-d}
\end{eqnarray}

Now, we rewrite these identities in terms of the Miwa shifts (both positive 
and negative) and demonstrate that they coincide with the AL equations. 
As in section \ref{sec-miwa}, we put 
\begin{equation}
  \kappa = 0.
\end{equation}

First let us consider the positive subhierarchy. 
The definition of the positive Miwa shifts, \eref{def-miwa-pos}, can be 
formulated for the $n$-dependent AL tau-functions as 
\begin{equation}
  \myShifted[E]{\xi} \omega_{n} 
  = 
  \myShifted{\xi} \omega_{n-1},
  \qquad
  \omega_{n} = \tau_{n}, \sigma_{n} \; \mbox{or}\; \rho_{n}.
\end{equation}
Replacing in \eref{alh-nz-a} and \eref{alh-nz-d} $\zeta$ with $\xi$ and 
expressing, using the preceding equation, $\myShift{\zeta}$ in terms of 
$\myShift[E]{\xi}$, one can immediately obtain that the resulting equations 
are nothing but \eref{alh-pos-b} and \eref{alh-pos-a} (up to the trivial shift 
$n \to n+1$). 

One can consider the case of negative flows in a similar way. 
Using the definition of the negative Miwa shifts, \eref{def-miwa-neg}, 
which we rewrite as 
\begin{equation}
  \begin{array}{lcl}
  \myshifted{\zeta}{\tau_{n}} 
  & = & 
  \myBarShift[E]{\eta}{\tau_{n}}, 
  \\ 
  \myshifted{\zeta}{\sigma_{n}}  
  & = & 
  - 
  \eta \,
  \myBarShift[E]{\eta} \sigma_{n} 
  \end{array}
\end{equation}
with $\eta = \zeta^{-1}$, 
one can easily verify that equations \eref{alh-nz-b} and \eref{alh-nz-c} 
turn out to be the negative AL equations \eref{alh-neg-a} and 
\eref{alh-neg-b} correspondingly.

We do not present here a verification of the $\rho$-equations, 
\eref{alh-pos-c} and \eref{alh-neg-c}, 
because it is a straightforward repetition of the preceding  
consideration that can be done using, instead of \eref{alh-fay-a} and 
\eref{alh-fay-c}, similar identities that stem from \eref{FayR} and \eref{FayBiR}.  
To summarize, we have shown that the tau-functions defined by 
\eref{def-tau} and \eref{def-sigmarho} solve the AL equations. 

Now, after establishing the structure of the solutions 
we have only to write down the dependence of $\langle a |$ and $\langle b |$ 
(and hence of $\mymatrix{A}$- and $\mymatrix{B}$-matrices) on 
$n$, $\mathrm{z}$- and $\mathrm{\bar{z}}$-variables. 
This means that we have to introduce $n$-, $\mathrm{z}$- and 
$\mathrm{\bar{z}}$-dependence in the way that ensures that the translation 
$n \to n+1$ coincides with the application of $\myShift{0}$-operator while 
the action of the Miwa shifts defined by \eref{alh-miwa} leads to 
\eref{miwa-mat-pos} and \eref{miwa-mat-neg}. 
This can be easily done by introducing the function 
$\myf_{n}(\mathrm{z},\mathrm{\bar{z}}; h)$, 
\begin{equation}
  \myf_{n}(\mathrm{z},\mathrm{\bar{z}}; h) 
  = 
  n \ln h 
  + 
  i\sum_{k=1}^{\infty} \left( h^{-k} z_{k} + h^{k} \bar{z}_{k} \right), 
\label{alh-sol-f}
\end{equation}
which satisfies 
$ e^{\myf_{n+1}(\mathrm{z},\mathrm{\bar{z}}; h)} 
  = 
  h \, e^{\myf_{n}(\mathrm{z},\mathrm{\bar{z}}; h)} 
$ 
and
\begin{eqnarray}
  \myShift[E]{\xi} e^{\myf_{n}(\mathrm{z},\mathrm{\bar{z}}; h)} 
  & = & 
  \left( 1 - \xi / h \right) 
  e^{\myf_{n}(\mathrm{z},\mathrm{\bar{z}}; h)}, 
\label{alh-f-pos} 
\\
  \myBarShift[E]{\eta} e^{\myf_{n}(\mathrm{z},\mathrm{\bar{z}}; h)} 
  & = & 
  \left( 1 - \eta h \right) 
  e^{\myf_{n}(\mathrm{z},\mathrm{\bar{z}}; h)}. 
\label{alh-f-neg} 
\end{eqnarray}
It is easy to verify that the choice 
\begin{equation}
  \begin{array}{lclcl}
  \langle a | 
  & = & 
  \langle a_{n}(\mathrm{z},\mathrm{\bar{z}}) | 
  & = & 
  \langle a^{(0)} | 
  e^{ - \myf_{n}(\mathrm{z},\mathrm{\bar{z}}; \mymatrix{R})}, 
  \\
  \langle b | 
  & = & 
  \langle b_{n}(\mathrm{z},\mathrm{\bar{z}}) | 
  & = & 
  \langle b^{(0)} | 
  e^{ \myf_{n}(\mathrm{z},\mathrm{\bar{z}}; \mymatrix{L})} 
  \end{array} 
\end{equation}
with constant $\langle a^{(0)} |$ and $\langle b^{(0)} |$ 
(which, in its turn, determines the dependence of $\mymatrix{A}$ and 
$\mymatrix{B}$ on $n$, $\mathrm{z}$ and $\mathrm{\bar{z}}$), 
leads to the proper translation properties,
$ \langle a_{n+1} | = \langle a_{n} | \, \mymatrix{R}^{-1}$ 
and
$ \langle b_{n+1} | = \langle b_{n} | \, \mymatrix{L}$ 
(and, hence, to  
$ \mymatrix{A}_{n+1} = \mymatrix{A}_{n} \mymatrix{R}^{-1}$ 
and 
$ \mymatrix{B}_{n+1} = \mymatrix{B}_{n} \mymatrix{L}$)
as well as the fulfilment of \eref{miwa-mat-pos} and \eref{miwa-mat-neg}. 

Finally, one has to count the constants that appear in the preceding formulae. 
At present, we have, apparently, $4N$ constants 
(except of the elements of the matrices $\mymatrix{L}$ and $\mymatrix{R}$): 
the components of the columns $| \,\alpha\, \rangle$ and $| \,\beta\, \rangle$ 
and of the rows $\langle a^{(0)} |$ and $\langle b^{(0)} |$. 
A simple analysis leads to the conclusion that these constants are actually 
combined (through products and ratios) in $2N$ ones. 
So, we can fix from the beginning, for example, $\langle a^{(0)} |$ and 
$\langle b^{(0)} |$ with $| \,\alpha\, \rangle$ and $| \,\beta\, \rangle$ 
playnig the role of the arbitrary parameters of the solution. 
In what follows we take 
\begin{equation}
  \langle a^{(0)} | = \langle 1 |, 
  \qquad
  \langle b^{(0)} | = \langle 1 | \mymatrix{L}
\end{equation}
where $\langle 1 |$ is the row with all components equal to $1$ 
and use instead of $| \,\beta\, \rangle$ the new constant vector 
\begin{equation}
  | \,\gamma\, \rangle = \mymatrix{R}^{-1} | \,\beta\, \rangle 
\end{equation}
(the advantages of this choice will be revealed next, when we discuss the 
`physical' involution).

Now we can write 
\begin{eqnarray} 
  \mymatrix{A}_{n}(\mathrm{z},\mathrm{\bar{z}}) 
  = 
  \left(
    \frac{ \alpha_{j} e^{-\myf_{n}(\mathrm{z},\mathrm{\bar{z}}; R_{k})} }
         { L_{j} - R_{k} }
  \right)_{j,k=1, ..., N}, 
\label{alh-sol-A} 
\\  
  \mymatrix{B}_{n}(\mathrm{z},\mathrm{\bar{z}}) 
  = 
  \left(
    \frac{ \gamma_{j} e^{\myf_{n}(\mathrm{z},\mathrm{\bar{z}}; L_{k})} }
         { L_{k}^{-1} - R_{j}^{-1} }
  \right)_{j,k=1, ..., N} 
\label{alh-sol-B} 
\end{eqnarray}
and formulate the main result of this section.

\begin{proposition} \label{prop-alh}
The $N$-soliton solution for equations 
\eref{alh-fr-pos-q}--\eref{alh-fr-neg-r} 
decribing the ALH is given by 
\begin{eqnarray}
  q_{n}(\mathrm{z},\mathrm{\bar{z}}) 
  & = & 
  \sum_{j,k=1}^{N} 
  e^{-\myf_{n}(\mathrm{z},\mathrm{\bar{z}}; R_{j})} 
  \mymatrix{F}^{(jk)}_{n}(\mathrm{z},\mathrm{\bar{z}}) 
  \gamma_{k}, 
  \\ 
  r_{n}(\mathrm{z},\mathrm{\bar{z}}) 
  & = & 
  \sum_{j,k=1}^{N} 
  e^{ \myf_{n}(\mathrm{z},\mathrm{\bar{z}}; L_{j})} 
  \mymatrix{G}^{(jk)}_{n}(\mathrm{z},\mathrm{\bar{z}}) 
  \alpha_{k} 
\end{eqnarray}
where 
$\mymatrix{M}^{(jk)}$ stands for the $jk^{\mathrm{th}}$ entry of a matrix 
$\mymatrix{M}$ and the matrices $\mymatrix{F}_{n}$ and $\mymatrix{G}_{n}$ are 
defined as  
\begin{eqnarray} 
  \mymatrix{F}_{n}(\mathrm{z},\mathrm{\bar{z}}) 
  & = & 
  \left[
    \mymatrix{1} 
    + 
    \mymatrix{B}_{n}(\mathrm{z},\mathrm{\bar{z}}) 
    \mymatrix{A}_{n}(\mathrm{z},\mathrm{\bar{z}}) 
  \right]^{-1}, 
\\
  \mymatrix{G}_{n}(\mathrm{z},\mathrm{\bar{z}}) 
  & = & 
  \left[
    \mymatrix{1} 
    + 
    \mymatrix{A}_{n}(\mathrm{z},\mathrm{\bar{z}}) 
    \mymatrix{B}_{n}(\mathrm{z},\mathrm{\bar{z}}) 
  \right]^{-1}. 
\end{eqnarray} 
Here, the matrices 
$\mymatrix{A}_{n}(\mathrm{z},\mathrm{\bar{z}})$, 
$\mymatrix{B}_{n}(\mathrm{z},\mathrm{\bar{z}})$ 
and the function $\myf_{n}(\mathrm{z},\mathrm{\bar{z}}; h)$  
are given by \eref{alh-sol-A}, \eref{alh-sol-B} and \eref{alh-sol-f},  
while $\alpha_{k}$ and $\gamma_{k}$ ($k=1,...,N$) are arbitrary constants. 

\end{proposition} 
Note that we have derived the $N$-soliton solutions for the ALH using a 
minimal amount of calculations: after rewriting several times the 
Fay identities the only calculation we made was to summarize an elementary 
series to obtain/verify properties \eref{alh-f-pos} and \eref{alh-f-neg} 
of $\myf_{n}(\mathrm{z},\mathrm{\bar{z}})$. 

\medskip 

Now we are to discuss the question which was not raised before, 
namely the one related to the involution: what restriction one has to impose 
on soliton matrices to meet the `physical' condition for the bright solitons,
\begin{equation}
  r_{n} = - q_{n}^{*}, 
\label{involution}
\end{equation} 
where $(...)^{*}$ indicates the complex conjugation. 
In other words, we are going to present here solutions for the `physical' 
version of the AL system: 
\begin{eqnarray}
  ( \myShift[E]{\xi} - 1 ) q_{n} 
  & = & 
  \xi 
  \left[ 1 + (\myShifted[E]{\xi}{q_{n}})q_{n}^{*} \right] 
  \myShifted[E]{\xi}{q_{n+1}}, 
\label{alhi-fr-pos}
\\
  ( \myBarShift[E]{\eta} - 1 ) q_{n} 
  & = & 
  \eta 
  \left[ 1 + (\myBarShifted[E]{\eta}{q_{n}})q_{n}^{*} \right] 
  \myBarShifted[E]{\eta}{q_{n-1}}. 
\label{alhi-fr-neg}
\end{eqnarray}
It can be easily verified, starting from proposition \ref{prop-alh}, 
 that to meet \eref{involution} one has to impose 
the two restrictions: 
\begin{equation}
  \mymatrix{L} 
  = 
  \left( \mymatrix{R}^{*} \right)^{-1} 
\label{alh-inv-LR}
\end{equation} 
and 
\begin{equation}
  \alpha_{j} 
  = 
  - \gamma_{j}^{*}. 
\end{equation} 
Indeed, the function $\myf_{n}(\mathrm{z},\mathrm{\bar{z}}; h)$ has the 
property 
\begin{equation}
  \myf_{n}\left(\mathrm{z},\mathrm{\bar{z}}; 1 / h^{*} \right)
  = 
  - \myf^{*}_{n}\left(\mathrm{z},\mathrm{\bar{z}}; h \right) 
\end{equation}
(provided we identify $\bar{z}_{j}$ and $z_{j}^{*}$) 
which, together with \eref{alh-inv-LR}, implies that 
\begin{equation}
  \mymatrix{B}_{n}(\mathrm{z},\mathrm{\bar{z}}) 
  = 
    \mymatrix{A}^{*}_{n}(\mathrm{z},\mathrm{\bar{z}}) 
\end{equation}
and hence 
$
  \mymatrix{G}_{n}(\mathrm{z},\mathrm{\bar{z}}) 
  = 
  \mymatrix{F}^{*}_{n}(\mathrm{z},\mathrm{\bar{z}}) 
$, 
leading to 

\begin{proposition} \label{prop-alhi}
The $N$-soliton solution for equations 
\eref{alhi-fr-pos}, \eref{alhi-fr-neg} 
decribing the ALH with the involution \eref{involution}
is given by 
\begin{equation}
  q_{n}(\mathrm{z},\mathrm{\bar{z}}) 
  = 
  - \sum_{j,k=1}^{N} 
  e^{-\myf_{n}(\mathrm{z},\mathrm{\bar{z}}; R_{j})} 
  \mymatrix{F}^{(jk)}_{n}(\mathrm{z},\mathrm{\bar{z}}) 
  \alpha^{*}_{k} 
\end{equation}
where the matrix $\mymatrix{F}_{n}$ is defined as  
\begin{equation} 
  \mymatrix{F}_{n}(\mathrm{z},\mathrm{\bar{z}}) 
  = 
  \left[
    \mymatrix{1} 
    + 
    \mymatrix{A}^{*}_{n}(\mathrm{z},\mathrm{\bar{z}}) 
    \mymatrix{A}_{n}(\mathrm{z},\mathrm{\bar{z}}) 
  \right]^{-1} 
\end{equation} 
with 
\begin{equation} 
  \mymatrix{A}_{n}(\mathrm{z},\mathrm{\bar{z}}) 
  = 
  \left(
    \frac{ \alpha_{j} R_{j}^{*} 
           e^{-\myf_{n}(\mathrm{z},\mathrm{\bar{z}}; R_{k})} }
         { 1 - R_{j}^{*} R_{k} }
  \right)_{j,k=1, ..., N}
\end{equation}
Here, the function $\myf_{n}(\mathrm{z},\mathrm{\bar{z}}; h)$  
is given by \eref{alh-sol-f},  
while $\alpha_{k}$ ($k=1,...,N$) are arbitrary constants. 

\end{proposition} 

This concludes the derivation of the $N$-bright soliton solutions for the ALH. 
It should be noted that it is a slight exaggeration to call the preceding 
procedure `derivation': the only 
`guesswork' one has to do is to assert \eref{alh-n}. All the rest is simple  
rewriting the identities we have already obtained. 
So, the preceding consideration 
can be viewed as an example of usefulness of the Fay identities as a tool for 
obtaining solutions for integrable systems by means of elementary calculations.

\ack 

We would like to thank the referees for careful reading the manuscript and for 
providing useful comments and suggestions that helped us to improve the paper.


\appendix

\section{Proof of \eref{shifted-sigma}. \label{proof-A}}
The aim of this and the following sections is to prove some of the identities 
that were used in section \ref{sec-mat}. 

Recalling definition \eref{def-FG} 
and using \eref{eq-Sylvester} one can obtain
\begin{eqnarray}
  \myshifted{\zeta}{\mymatrix{F}}^{-1} 
  & = & 
  \mymatrix{1} 
  + 
  \mymatrix{B} 
  \left( \mymatrix{L} - \zeta \right)
  \mymatrix{A} 
  \left( \mymatrix{R} - \zeta \right)^{-1} 
\\  
  & = & 
  \mymatrix{1} 
  + 
  \mymatrix{B} 
  \left[ 
    \mymatrix{A} 
    \left( \mymatrix{R} - \zeta \right)
    + 
    | \,\alpha\, \rangle \langle a |
  \right] 
  \left( \mymatrix{R} - \zeta \right)^{-1} 
\\  
  & = & 
  \mymatrix{F}^{-1}  
  + 
  \mymatrix{B} 
  | \,\alpha\, \rangle \langle \myshifted{\zeta}{a} |. 
\end{eqnarray}
After multiplying by $\myshifted{\zeta}{\mymatrix{F}}$ from the left 
and by $\mymatrix{F}$ from the right, the last equation becomes
\begin{equation}
  \myshifted{\zeta}{\mymatrix{F}} 
  = 
  \mymatrix{F} 
  - 
  \myshifted{\zeta}{\mymatrix{F}}
  \mymatrix{B} 
  | \,\alpha\, \rangle \langle \myshifted{\zeta}{a} | \mymatrix{F}
\end{equation} 
and, 
after multiplying by $\langle \myshifted{\zeta}{a} | $ from the left 
and by $| \,\beta\, \rangle $ from the right, 
\begin{equation}
  \langle \myshifted{\zeta}{a} | 
  \myshifted{\zeta}{\mymatrix{F}} 
  | \,\beta\, \rangle 
  = 
  \left(
    1 
    - 
    \langle \myshifted{\zeta}{a} | 
    \myshifted{\zeta}{\mymatrix{F}} \mymatrix{B} 
    | \,\alpha\, \rangle 
  \right)
  \langle \myshifted{\zeta}{a} | 
  \mymatrix{F} 
  | \,\beta\, \rangle. 
\end{equation} 
Using one of identities \eref{shifted-tau-n} 
(shifted by $\myShift{\zeta}$)  
\begin{equation}
  \tau / \myshifted{\zeta}{\tau} 
  = 
  1 - 
  \langle 
    \myshifted{\zeta}{a} | \myshifted{\zeta}{\mymatrix{F}} \mymatrix{B} 
  | \alpha \rangle, 
\end{equation}
and the definition of $\sigma$ \eref{def-sigmarho} one arrives at the first 
of the identities \eref{shifted-sigma}. 

In a similar way, one can write 
\begin{eqnarray}
  \myshifted{\zeta}{\mymatrix{F}}^{-1} 
  & = & 
  \mymatrix{1} 
  + 
  \mymatrix{B} 
  \left( \mymatrix{L} - \zeta \right)
  \myshifted{\zeta}{\mymatrix{A}} 
\\  
  & = & 
  \mymatrix{1} 
  + 
  \left[ 
    \left( \mymatrix{R} - \zeta \right)
    \mymatrix{B} 
    - 
    | \,\beta\, \rangle \langle b |
  \right] 
  \myshifted{\zeta}{\mymatrix{A}} 
\\  
  & = & 
  \left( \mymatrix{R} - \zeta \right) 
  \mymatrix{F}^{-1}  
  \left( \mymatrix{R} - \zeta \right)^{-1}
  - 
  | \,\beta\, \rangle \langle b |
  \myshifted{\zeta}{\mymatrix{A}} 
\end{eqnarray}
which leads to 
\begin{equation}
  \left( \mymatrix{R} - \zeta \right)^{-1} 
  \myshifted{\zeta}{\mymatrix{F}} 
  = 
  \mymatrix{F} 
  \left( \mymatrix{R} - \zeta \right)^{-1} 
  + 
  \mymatrix{F} 
  \left( \mymatrix{R} - \zeta \right)^{-1} 
  | \,\beta\, \rangle \langle b |
  \myshifted{\zeta}{\mymatrix{A}} 
  \myshifted{\zeta}{\mymatrix{F}} 
\end{equation}
and hence to 
\begin{equation}
  \left( \mymatrix{R} - \zeta \right)^{-1} 
  \myshifted{\zeta}{\mymatrix{F}} | \,\beta\, \rangle 
  = 
  \left( 
    1
    + 
    \langle b |
    \myshifted{\zeta}{\mymatrix{A}} 
    \myshifted{\zeta}{\mymatrix{F}} 
    | \,\beta\, \rangle 
  \right) 
  \mymatrix{F} 
  \left( \mymatrix{R} - \zeta \right)^{-1} 
  | \,\beta\, \rangle. 
\end{equation}
Noting that $\mymatrix{A}\mymatrix{F} = \mymatrix{G}\mymatrix{A}$ and 
using the shifted second identity \eref{shifted-tau-n}, 
\begin{equation}
  \tau / \myshifted{\zeta}{\tau} 
  = 
  1 + 
  \langle b | 
    \myshifted{\zeta}{\mymatrix{A}} 
    \myshifted{\zeta}{\mymatrix{F}} 
  | \beta \rangle, 
\end{equation}
one arrives, 
after multiplication by 
$
  \langle a | = 
  \langle \myshifted{\zeta}{a} | \left( \mymatrix{R} - \zeta \right), 
$
at the second of the 
identities \eref{shifted-sigma}.

\section{Proof of \eref{innerST}. \label{proof-B}}

To prove \eref{innerST}, let us consider the action of the matrix 
$\mathcal{K}_{\sigma\tau}(\zeta,\emptyset)$ on the vector 
$\mymatrix{F}| \beta \rangle$. 
The following simple calculations 
\begin{eqnarray}
  \left( \mymatrix{R} - \zeta \right) 
  \mathcal{K}_{\sigma\tau}(\zeta,\emptyset)
  & = & 
  \mymatrix{1} 
  + 
  \left(\mymatrix{R} - \zeta\right)
  \mymatrix{B} 
  \left(\mymatrix{L} - \zeta\right)^{-1}
  \mymatrix{A} 
\\
  & = & 
  \mymatrix{1} 
  + 
  \left[ 
    \mymatrix{B} 
    \left(\mymatrix{L} - \zeta\right) 
    + 
    | \,\beta\, \rangle \langle b |
  \right]   
  \left(\mymatrix{L} - \zeta\right)^{-1}
  \mymatrix{A} 
\\
  & = & 
  \mymatrix{1} 
  + 
  \mymatrix{B} 
  \mymatrix{A} 
  + 
  | \,\beta\, \rangle \langle \myshifted{\bar\zeta}{b} |
  \mymatrix{A} 
\end{eqnarray}
lead to 
\begin{eqnarray}
  \left( \mymatrix{R} - \zeta \right) 
  \mathcal{K}_{\sigma\tau}(\zeta,\emptyset)
  \mymatrix{F} 
  | \,\beta\, \rangle 
  & = & 
  \left( 
    \mymatrix{1} 
    + 
    | \,\beta\, \rangle \langle \myshifted{\bar\zeta}{b} |
    \mymatrix{A} 
    \mymatrix{F} 
  \right)
  | \,\beta\, \rangle 
\\ & = & 
  \left( 
    1 
    + 
    \langle \myshifted{\bar\zeta}{b} |
    \mymatrix{A} 
    \mymatrix{F} 
    | \,\beta\, \rangle 
  \right)
  | \,\beta\, \rangle 
\\ & = & 
  \frac{ \myshifted{\bar\zeta}{\tau} }{ \tau } \; 
  | \,\beta\, \rangle. 
\label{proof-B-a}
\end{eqnarray}
Noting that 
\begin{equation}
  \mathcal{K}_{\sigma\tau}(\zeta,\myset{Y}) 
  = 
  \myShift{ \myset{Y} }
  \mathcal{K}_{\sigma\tau}(\zeta,\emptyset), 
\label{proof-B-b}
\end{equation}
presenting the left-hand side of \eref{innerST} as 
\begin{equation}
  \frac{ \myshifted{\myset{X}\zeta}{\sigma} 
         \myshifted{\myset{Y}\bar\zeta}{\tau} }
       { \myshifted{\myset{X}}{\tau} 
         \myshifted{\myset{Y}}{\tau} }
  =  
  \langle \myshifted{\myset{X}}{a} | 
  \myshifted{\myset{X}}{\mymatrix{F}}
  \left(\mymatrix{R} - \zeta\right)^{-1}
  | \beta \rangle  
  \frac{ \myshifted{\myset{Y}\bar\zeta}{\tau} }
       { \myshifted{\myset{Y}}{\tau} },
\end{equation} 
(we use the second of the identities \eref{shifted-sigma}) 
and replacing 
$ | \beta \rangle  
  \myshifted{\myset{Y}\bar\zeta}{\tau} / 
  \myshifted{\myset{Y}}{\tau} 
$
with the expression that follows from \eref{proof-B-a} and \eref{proof-B-b}, 
\begin{equation}
  | \beta \rangle  
  \frac{ \myshifted{\myset{Y}\bar\zeta}{\tau} }
       { \myshifted{\myset{Y}}{\tau} }
  = 
  \left(\mymatrix{R} - \zeta\right) 
  \mathcal{K}_{\sigma\tau}(\zeta,\myset{Y})
  \myshifted{\myset{Y}}{\mymatrix{F}}  
  | \beta \rangle,  
\end{equation}
one concludes the proof of the identity \eref{innerST}.


\end{document}